\documentclass[12pt]{article}
\usepackage{graphicx}
\begin{document}
%
%
\begin{quote}
\raggedright To be submitted to:
\end{quote}
\vspace{-2.2cm}
\begin{center}
{\large Nucl. Phys A.}
\end{center}
%
%
\begin{center}
{\Large The $\pi\pi$ interaction in nuclear matter from a study 
        of the $\pi^+ A \rightarrow \pi^+ \pi^{\pm} A'$ reactions}
\end{center}
%
%
\begin{center}
\normalsize
F. Bonutti$^{a,b}$, P. Camerini$^{a,b}$, E. Fragiacomo$^{a,b}$, 
N. Grion$^{a,}$\footnote[1]
{Corresponding author, electronic mail: Nevio.Grion@ts.INFN.it}, 
R. Rui$^{a,b}$, J.T. Brack$^{c,}$\footnote[2] 
{Present address: University of Colorado, Boulder CO 80309-0446, USA},
L. Felawka$^c$, E.F. Gibson$^f$, G. Hofman$^{d,2}$, M. Kermani$^{d,}$, 
E.L. Mathie$^e$, R. Meier$^{c,}$\footnote[3]
{Permanent address: Physikalisches Institut, Universit\"{a}t
T\"{u}bingen, 72076 T\"{u}bingen, Germany}, 
 D. Ottewell$^c$, K. Raywood$^c$, M.E. Sevior$^g$, G.R. Smith$^c$ 
and R. Tacik$^e$.
\end{center}
%
%

\begin{center}
\small{\it 
$^a$ Istituto Nazionale di Fisica Nucleare, 34127 Trieste, Italy \\
$^b$ Dipartimento di Fisica dell'Universita' di Trieste, 34127 Trieste, 
     Italy\\
$^c$ TRIUMF, Vancouver, B.C., Canada V6T 2A3 \\
$^d$ Department of Physics and Astronomy, University of British Columbia,
     Vancouver, B.C., Canada V6T 2A6 \\
$^e$ University of Regina, Regina, Saskatchewan, Canada S4S 0A2 \\
$^f$ California State University, Sacramento CA 95819, USA \\
$^g$ School of Physics, University of Melbourne, Parkville, Vic., 3052,
     Australia
} 
\end{center}                   
\begin{center}
\large{The CHAOS Collaboration} 
\end{center}

%
%
\setlength{\baselineskip}{2.3ex}         
{\small
The pion-production reactions $\pi^+ A \rightarrow \pi^+\pi^{\pm} A'$ 
were studied on $^{2}H$, $^{12}C$, $^{40}Ca$, and $^{208}Pb$ nuclei at 
an incident pion energy of $T_{\pi^{+}}$=283 MeV. Pions were detected 
in coincidence using the CHAOS spectrometer. The experimental results 
are reduced to differential cross sections and compared to both 
theoretical predictions and the reaction phase space. The composite ratio 
$\cal C$$_{\pi\pi}^A$ between the $\pi^{+}\pi^{\pm}$ invariant masses 
on nuclei and on the nucleon is also presented. Near the $2m_{\pi}$ 
threshold pion pairs couple to $(\pi\pi)_{I=J=0}$ when produced in the 
$\pi^+\rightarrow \pi^+\pi^-$ reaction channel.  There is a marked 
near-threshold enhancement of $\cal C$$_{\pi^+\pi^-}^A$ which is 
consistent with theoretical predictions addressing the partial restoration 
of chiral symmetry in nuclear matter. Furthermore, the behaviour of 
$\cal C$$_{\pi^+\pi^-}^A$ is well described when the restoration of 
chiral symmetry is combined with standard P-wave renormalization of pions 
in nuclear matter. On the other hand, nuclear matter only weakly influences 
$\cal C$$_{\pi^+\pi^+}^A$, which displays a flat behaviour throughout the 
energy range regardless of $A$.
}

\newpage

\normalsize
%
%
{\bf 1. Introduction}

The influence of the nuclear medium on the $\pi\pi$ interaction was 
investigated at TRIUMF using the pion induced pion-production reactions
$\pi^+A \rightarrow \pi^+\pi^{\pm}A'$ (henceforth labelled $\pi 2\pi$). 
The initial study was directed to the reactions in deuterium, that is, 
to $\pi^+ n(p) \rightarrow \pi^+\pi^- p(p)$ and 
$\pi^+ p(n) \rightarrow \pi^+\pi^+ n(n)$, in order to understand the 
$\pi 2\pi$ behaviour on both a neutron and a proton through quasi-free 
reactions. The $\pi 2\pi$ process was then examined on the complex nuclei 
$^{12}C$, $^{40}Ca$ and $^{208}Pb$ in order to study possible $\pi\pi$ 
medium modifications by direct comparison of the $\pi 2\pi$ data.  To
ensure the validity of this approach, the following experimental method
was applied.  All the $\pi 2\pi$ data were taken under the same kinematical 
conditions.  The final $\pi^+\pi^{\pm}$ pairs were detected in coincidence 
to ensure a reliable identification of $\pi 2\pi$ 
events.  Pion pairs were analysed down to $0^\circ$ $\pi\pi$ opening 
angles to determine the $\pi\pi$ invariant mass at the $2m_{\pi}$ threshold. 
The energy of the incident pion beam, $T_{\pi^+}$=283 MeV, was 
chosen to enable the investigation of the $\pi A\rightarrow\pi\pi A'$ 
reaction in the near-threshold region.

In early experimental works on $\pi 2\pi$ in nuclei it was observed
that (i) pions preferentially populate the low-energy part 
of the kinetic energy spectra \cite{Grion:one,Grion:two,
Camerini:one}, and (ii) the $\pi^+\pi^-$ invariant mass 
$M_{\pi^+\pi^-}^A$ peaks increasingly toward the $2m_\pi$ 
threshold as the nucleus mass number increases
\cite{Camerini:one}. The property (i) could not be explained
in terms of the $\pi^+A \rightarrow \pi^+\pi^- p[A-1]$ phase 
space \cite{Grion:one}.  Nor was a detailed model of the 
$A(\pi^+,\pi^+\pi^-$) reaction \cite{Oset:one} able 
to reproduce the low-energy pion yield, although it embodied 
the kinematical limits of the experimental apparatus. The 
same model, however, could correctly predict the total cross 
section of the $\pi 2\pi$ reaction in nuclei \cite{Bonutti:zero},
as well as many-fold differential cross sections \cite{Grion:two}. 
A novel approach \cite{Schuck:one} was then employed to explain the 
near threshold behaviour of $M_{\pi^+\pi^-}^A$ , i.e. the property (ii).  
In this approach a $\pi\pi$ pair is considered as a strongly 
interacting system when the two pions couple to the I=J=0 quantum 
numbers. The $(\pi\pi)_{I=J=0}$ properties in nuclear matter 
are studied by dressing the single-pion propagator 
to account for the $P-$wave coupling of pions to $particle-hole$  
and $\Delta-hole$ configurations. The model is able to explain the 
general features of the $M_{\pi^+\pi^-}^A$ distributions, which 
are predicted to increasingly accumulate strength near the 
$2m_{\pi}$ threshold for $\rho$, the nuclear medium density, 
approaching $\rho_0$, the saturation density. However, within the same 
theoretical framework, the near-threshold strength 
of the $\pi\pi$ $T-$matrix is considerably reduced when the 
$\pi\pi$ interaction is constrained to be chiral symmetric
\cite{Aouissat:one}.  This may indicate that effects other 
than the in-medium $(\pi\pi)_{I=J=0}$ interaction contribute 
to the observed strength. Conversely, the absence of any 
in-medium modification of the $(\pi\pi)_{I=J=0}$ interaction 
leads to $M_{\pi^+\pi^-}^A$ distributions which lack of 
strength at threshold.  This is shown in the work of 
\cite{Camerini:one}, which compares the experimental results 
with the model predictions of \cite{Oset:one}.

Some $\pi 2\pi$ articles were recently published by the 
CHAOS collaboration at TRIUMF. The novel data highlighted 
some properties of the in-vacuum $\pi\pi$ interaction
\cite{Bonutti:three,Moh:one,Moh:two}, as well as the 
in-medium modifications of the $\pi\pi$ interaction
\cite{Bonutti:one,Bonutti:two,Bonutti:four}. The appearance 
of the CHAOS results have renewed theoretical interest, 
which now bases the interpretation of the $\pi 2\pi$ 
and $\pi\pi$ data on some common features: 
\begin{enumerate}
  \item The $\pi\pi$ interaction is strongly influenced and modified
        by the presence of the nuclear medium when the $\pi\pi$ 
        interaction occurs in the $(\pi\pi)_{I=J=0}$ channel, 
        conventionally called the $\sigma-$channel
        \cite{Schuck:one,Schuck:two,Chiang:one,Vicente:one,Hatsuda:one}. 
  \item Nuclear matter affects the $(\pi\pi)_{I=2,J=0}$ 
        interaction only weakly \cite{Vicente:one,Rapp:one}. 
  \item Models which only include standard many-body correlations, i.e. 
        the $P-$wave coupling of $\pi's$ to $p-h$ and $\Delta-h$
        configurations, are able to explain part of the observed 
        $M_{\pi^+\pi^-}^A$ yield near the $2m_{\pi}$ threshold
        \cite{Schuck:two,Vicente:one,Rapp:one}. 
\end{enumerate}
In recent theoretical works on the $(\pi\pi)_{I=J=0}$ interaction 
in nuclear matter \cite{Aouissat:two}, the effects of standard many-body 
correlations have been combined with those derived from the restoration of 
chiral symmetry in nuclear matter. The resulting $M_{\pi^+\pi^-}^A$ 
distributions regain strength near the $2m_{\pi}$ threshold
as $A$ (thus the average $\rho$) increases. Such behaviour was
outlined in some earlier theoretical works, which demonstrated that the 
$M_{\pi^+\pi^-}^A$ enhancement near threshold is a distinct consequence
of the partial restoration of the chiral symmetry at $\rho < \rho_0$ 
\cite{Hatsuda:one}. The purpose of this article is to present a 
comprehensive set of $\pi 2\pi$ data and to discuss them in the light 
of the most recent theoretical findings.

Some of the characteristics of the $\pi 2\pi$ process in nuclei were
presented in 
Refs.\cite{Grion:one,Grion:two,Camerini:one,Bonutti:zero,Bonutti:three,
Bonutti:one,Bonutti:two} and some observables, those most sensitive 
to the $\pi\pi$ interaction, were discussed in Ref.\cite{Bonutti:four}. 
These TRIUMF results are presented in the form of many-fold differential 
cross sections, which are of interest in the current work. Most other 
$\pi 2\pi$ measurements deal only with total cross sections. These include
old data taken with emulsion techniques \cite{oldpi2pidata:one}, and more 
recently, with a magnetic spectrometer similar to CHAOS \cite{Rahav:one}. 
Since $\pi 2\pi$ total cross sections are not part of the present discussion, 
these data will not be considered in this work. The article is organised 
as follows: some features of the experiment are reported in Sec. 2.
Sec. 3 deals with the data analysis. The available 
$\pi 2\pi$ models are presented in Sec. 4.
The $\pi 2\pi$ data are discussed in Sec. 5.
Other existing results are mentioned in Sec. 6.
Finally, conclusions are summarised in Sec. 7.

{\bf 2. The experiment}\\

The experiment was carried out at the TRIUMF Meson Facility. 
Incident pions were produced by the collision of 480 MeV 
protons on a 10 mm thick graphite target. The M11 pion beam 
line transported the 282.7 MeV pions to the final focus which 
was located at about 14 m from the production target. M11 
was tuned to focus the pion beam on the target, which was 
placed at the centre of the CHAOS spectrometer. Pion pairs 
were detected in coincidence to ensure a unique identification 
of the pion production process. \\

2.1 The pion beam, the beam monitor and the beam counting
     
The M11 pion channel was set to deliver positive pions with 
a central momentum p=398.5 MeV/c. The beam momentum spread 
($\Delta$p/p ) was defined by slits which were 
located at the intermediate dispersed focus of the channel. 
The particle beam was mainly composed of pions and protons, 
$\pi :p \approx 1:10$. The fraction of positive electrons and 
muons was negligible, typically $\pi :\mu :e=1.0:0.003:0.002$
\cite{TRIUMF:one}. The proton contamination was moderated by
a CH2 absorber placed at the mid-plane of the M11 channel: the 
energy deposited by protons in the absorber degraded the central 
momentum of the proton distribution so that most of them were 
intercepted by the slits of the channel. Pions were
finally discriminated from protons with a plastic scintillator 
counter placed transversally to the particle beam at about 
194 cm upstream of the target. The pulse height ($PH$) associated
with the passage of a proton through the plastic scintillator 
was over 3 times higher than the pulse height of a pion of the same 
momentum, i.e. $PH_p / PH_{\pi} \simeq 3$ at 400 MeV/c. This 
allowed for a fast discrimination between the two beam components 
and, ultimately, the determination of the beam composition, 
which was monitored throughout the experiment. The systematic 
uncertainty in determining the fraction of pion $f_{\pi}$ in 
the particle beam was about 1\%.

The beam monitor ($M_b$) consisted of 4 slabs of Pilot U 
plastic scintillator 3 mm thick, with a combined
cross-sectional area of 9$\times$26 cm$^2$. This area 
was wide enough to contain the full particle beam. Each 
segment of the monitor was designed to count roughly the 
same beam flux, i.e. 1.0-1.5 MHz for $^{12}C$,$^{40}Ca$ and 
$^{208}Pb$ targets, and $\le$0.5 MHz for $^2H$ (see Table 1). The 
four segments were coupled through lucite light guides to four 
fast-counting tubes in order to avoid piling-up of the output 
pulses.  The beam monitor was placed at about 161 cm from the 
CHAOS dipole tip edge to prevent any background arising from the 
interaction of pions and protons with the monitor medium from entering 
the spectrometer. However, with such an arrangement of the beam monitor,
an appreciable fraction of the pions crossing it decayed before 
reaching the target. The decay rate was $f_{d}=8\%$ and the 
uncertainty in assessing $f_{d}$ is negligible. The beam momentum 
spread ($\Delta$p/p at FWHM, [\%]), total intensity over all segments
($N$, [$10^6$particles/s]) and composition ($\pi:p:\mu:e$) at the 
monitor location are summarised in Table 1.  
\begin{table}[tbc]
\caption[Table]{
Characteristics of the particle beam at the monitor location: 
momentum spread ($\Delta$p/p at FWHM, [\%]), intensity ($N$, 
[$10^6$particles/s]) and composition \vspace{0.5cm}($\pi:p:\mu:e$). 
}
\begin{center}
  \begin{tabular}{lccc}  \hline \hline 
Nucleus                & $ \Delta$p/p & $N$ & $\pi : p : \mu : e $ \\ 
\hline 
$^{2}H$                       &   0.8 & 1.6 & 3.2:1.0:0.01:$<$0.01    \\ 
$^{12}C$,$^{40}Ca$,$^{208}Pb$ &   1.2 & 4-5 & 4.3:1.0:0.01:$<$0.01   \\ 
\hline\hline
  \end{tabular}
\end{center}
\end{table}

The beam monitor was only capable of counting the number of incident 
particle bursts $N_i$. Depending on its intensity, a $\sim$5ns FWHM 
beam burst may have contained one or more particles, i.e. pions and 
protons. Therefore the effective number of particles traversing the 
monitor is $N_m=\mu_p N_i$, where $\mu_p$ is the average number of 
particles per beam burst. Since particle counting follows Poisson 
statistics, the average number of particles is $\mu_p = -ln(1-r)/r$,
where $r=N_i/N_p$ and $N_p$ is the number of primary proton bursts per 
second 23.06$\times 10^6$. In the present measurement the maximum number 
of particles traversing a monitor single slab did not exceed 
$1.5\times 10^{6}$/s.  Thus, the percentage of particles not counted by 
the monitor was $100(\mu_p - 1)<3.5\%$, and the uncertainty associated 
with this correction is negligible.

The number of pions at the target location was finally given by the 
equation $N_{\pi} = \mu_p (1 - f_d) f_{\pi} N$, and the uncertainty in
assessing $N_{\pi}$ was slightly above 1\%.

2.2 The targets 

The targets used in this experiment were either solid self-supporting 
plates of carbon, calcium, and lead, or a vessel for the liquid 
deuterium. The targets were located at the centre of the dipole, and
their characteristics are summarised in Table 2.

The deuterium target consisted of a low-mass cylindrical vessel filled 
with liquid deuterium. It was continuously cooled by a refrigerator, and 
then warmed to its operating temperature, which was monitored both by an 
external helium gas thermometer and direct measurement of the vapour 
pressure of the deuterium. The variation of density due to the uncertainty 
in temperature (23.0$\pm$0.2K) was negligible. The variation of density due 
to the formation of bubbles throughout the liquid was larger, and most 
significant near the top of the target cell. The variation of thickness 
of the target cell due to bulging outward of the thin windows contributes 
a further uncertainty to the target thickness. Both of these effects were 
convoluted with measured beam profiles to determine the areal target 
thickness. The total number of scattering centres was evaluated by 
integrating the beam profile over that of the target, which was determined 
by tracing elastically scattered pions during a calibration run. 
The uncertainty in the areal target thickness was estimated to be 3\%. 
In order to evaluate the contribution of the background from the target 
vessel, a dedicated run was performed with the deuterium target emptied. 
Then, the rejection of events with the reaction vertex on the vessel 
reduced the background to about 2\%. A reconstructed interaction vertex 
for  a vessel of the same dimensions as the one used in the present 
experiment is shown in Fig. 2 of Ref.\cite{Moh:one}.

The solid targets were made of self-supporting plates surrounded
by a frame for handling.  The frame was made of lucite for the $^{12}C$ and
$^{208}Pb$ targets, and alluminum for the $^{40}Ca$ target. The target
thicknesses and areas are listed in Table 2. The areal size of the plates     
was large enough to entirely contain the beam spot. The 
beam envelope at the target location was measured previously
\cite{TRIUMF:two}, and was found to be entirely confined to an 
area of $\sim 40\times 40 mm^2$. This can be seen in Fig. 1,
\begin{figure}[htb]
 \centering
  \includegraphics*[angle=0,width=0.6\textwidth]{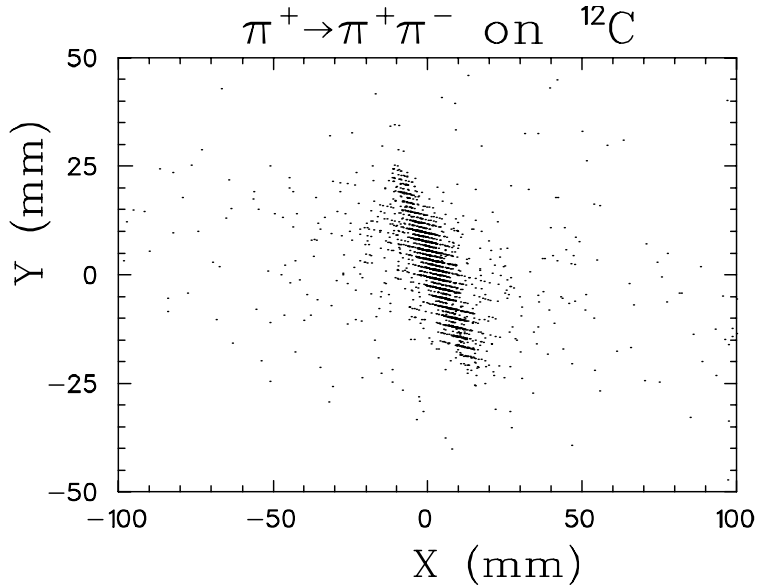}
  \setlength{\abovecaptionskip}{10pt}  
  \setlength{\belowcaptionskip}{15pt}  
  \caption{\footnotesize Horizontal beam profile at the $^{12}C$ target 
      location. The interaction vertex is reconstructed for 
      $\pi^+ \rightarrow \pi^+\pi^-$ events. The profile full-width of 
      the target about 55 mm is entirely contained in the physical 
      target, which has a width of 70 mm as reported in Table 2.}
\end{figure}
which illustrates the vertex reconstruction provided by the CHAOS 
spectrometer from $\pi^+ \rightarrow \pi^+\pi^-$ events on $^{12}C$. 
Analogous results were obtained for the other solid targets.
\begin{table}[tbc]
\caption[Table]{
Characteristics of the targets used in the experiment.
}
\begin{center}
  \begin{tabular}{ccccc}                      \hline \hline 
          &  $^{2}H$              &      $^{12}C$  &   $^{40}Ca$    &   $^{208}Pb$  \\ 
\hline
type      & liquid in cyl. vessel &  solid plate   &   solid plate  & solid plate    \\ 
thickness & H: 5 cm, $\phi$: 5 cm & 0.332 g/cm$^2$ & 0.180  g/cm$^2$&0.604 g/cm$^2$  \\ 
area      &         $-$        &$70\times61 mm^2$&$55\times46 mm^2$&$74\times58 mm^2$\\ 
\hline\hline
\end{tabular}
\end{center}
\end{table}

2.3 The CHAOS spectrometer and triggering system 

CHAOS is a magnetic spectrometer which was designed for the detection 
of multi-particle events in the medium-energy range \cite{CHAOS:one}. 
The magnetic field is generated by a dipole whose pole tip is 66 cm 
in diameter. The magnet is capable of producing a field intensity up 
to 1.6 T with an uniformity of about 1\%. There is a 12 cm 
bore at the centre for the insertion of targets. Fig. 2 illustrates 
\begin{figure}[h]
 \centering
  \includegraphics*[angle=90,width=0.7\textwidth]{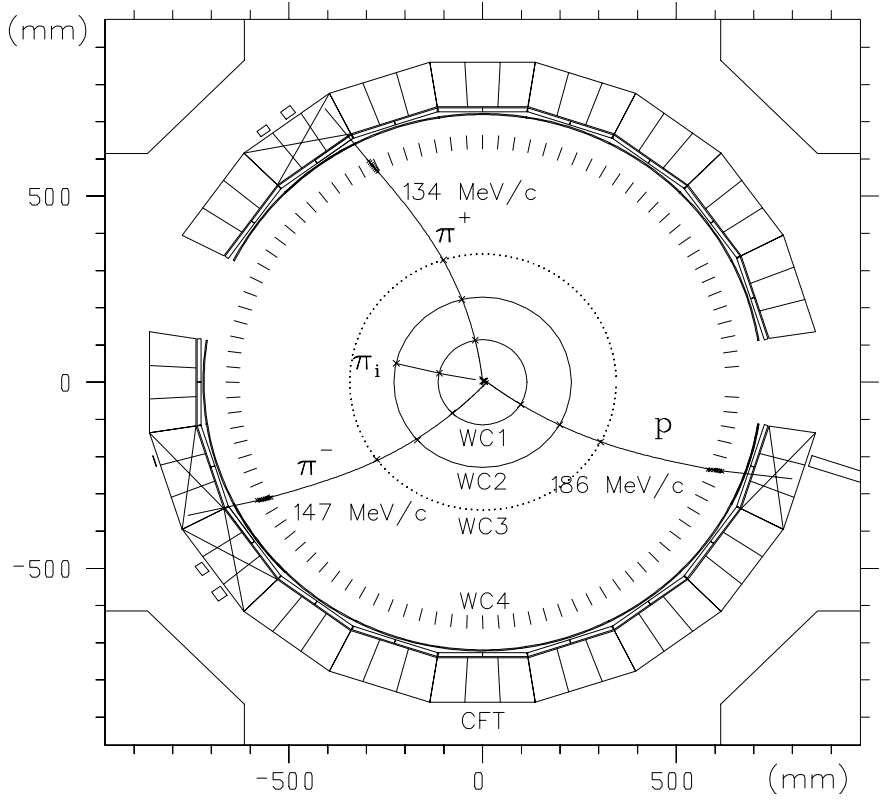}
  \setlength{\abovecaptionskip}{10pt}  
  \setlength{\belowcaptionskip}{15pt}  
  \caption{\footnotesize Reconstructed particle trajectories in CHAOS
    for the $\pi^+_i \rightarrow \pi^+\pi^-p$ reaction on $^{12}C$, 
    the geometrical disposition of the wire chambers (WC), the first 
    level trigger hardware (CFT) and the magnet return yokes in the 
    corners. Two CFT segments are removed to permit the particle beam 
    ($\pi_i$) to traverse the spectrometer. The CFT segments which are 
    hit by particles are marked with crosses, and the energy deposited
    in the first two layers ($\Delta E1$ and $\Delta E2$) is indicated 
    by boxes. The proton has a momentum  slightly above the CHAOS 
    threshold (185 MeV/c), and its energy is fully deposited in $\Delta E1$.}
\end{figure}
a reconstructed $\pi^+_i \rightarrow \pi^+\pi^-p$ event on $^{12}C$,
and shows the geometrical disposition of the wire chambers (WC), the CHAOS 
first level trigger hardware (CFT), and the magnet return yokes in the 
corners. WC1 and WC2 are multiwire proportional chambers which are 
capable of handling rates exceeding 5$\times 10^6$ particles/s for 
extended periods of time with high efficiency ($\sim$95\%). They are
cylindrical in shape, with diameters of 22.8 cm and 45.8 cm, respectively. 
WC3 is a cylindrical drift chamber designed to operate in a magnetic 
field \cite{CHAOS:two}.  Its diameter is 68.6 cm. The outermost 
chamber, WC4, is a vector drift chamber 122.6 cm in diameter, which 
operates in the tail of the magnetic field of CHAOS. The segments 
of WC3 and WC4 which were crossed by the incident particle beam were 
turned off. The CFT hardware consists of three adjacent cylindrical 
layers of fast-counting detectors \cite{CHAOS:three}. The first two 
layers ($\Delta E1$ and $\Delta E2$) are NE110 plastic 
scintillators 0.3 cm and 1.2 cm thick, respectively. $\Delta E1$ is 
72 cm from the magnet centre and spans a zenithal angle of 
$\pm 7^{\circ}$; thus, it defines the geometrical solid angle of CHAOS 
$\Omega$=1.5 sr. The third layer is a SF5 lead-glass 12.5 cm thick, 
about 5 radiation lengths, which is used as a Cerenkov counter. The 
three layers were segmented in order to provide an efficient triggering 
system for multi-particle events. Each segment covered an azimuthal 
angle (i.e., in-the-reaction plane) of 18$^\circ$. During data taking 
two segments were removed in order to allow the passage of the particle 
beam (see Fig. 2).

The experiment relied on two on-line triggers: the first ($1^{st}LT$) 
\cite{CHAOS:three} and second ($2^{nd}LT$) \cite{CHAOS:four} level 
trigger. The logical signals which were issued by the CFT following 
$\pi^+ \rightarrow \pi^+\pi^{\pm}$ events were initially pipe-lined 
and then filtered by requiring a coincidence $M_b\times 
(\Delta E1\times\Delta E2)_i\times (\Delta E1\times\Delta E2)_j$   
with $i,j=1,..,18$ and $i\neq j$. Such a logic analysis was performed 
at a rate of about 30 MHz.  Despite of the fast filtering, the 
$1^{st}LT$ trigger was overwhelmed ($>$99.99\%) by multi-particle 
events from competing reactions: quasi-elastic scattering (qes) 
$\pi^+\rightarrow\pi^+ p(d)$, and pion absorption (abs) 
$\pi^+ \rightarrow p p$. This called for a second and more 
selective triggering system, which was based on the information
delivered by the innermost wire chambers. For each event passed by  
the $1^{st}LT$, the $2^{nd}LT$ trigger calculated the momentum of each track, 
its polarity and the reaction vertex topology.  It then compared a combination 
of these parameters with predefined criteria. When the criteria were 
fulfilled the event was accepted and recorded on tape, otherwise a fast 
clear signal was sent to the readout electronics, and the two triggers were
reset. An entire $2^{nd}LT$ cycle was accomplished in $< 10\mu$s,
depending on the number of reconstructed tracks. The event rate recorded 
on tape ranged from 30 to 100 Hz depending on the nucleus studied.
Roughly 0.1\% of the recorded events were $\pi\rightarrow\pi\pi$ as 
determined by detailed off-line analysis.

2.4 The CHAOS acceptance and performance

In the present measurement CHAOS was operated with a magnetic field of
0.5 T. The field and the energy deposited by the particles emerging 
from the target in the WC's and $\Delta E1$ media determined the CHAOS 
threshold, which was 57 MeV/c for pions. The pion threshold can be 
observed in Fig. 3, which shows a diffusion plot of the $\pi^{\pm}$ 
momenta ($p_{\pi}$) versus their azimuthal angles ($\Theta_{\pi}$), for 
the $\pi^+ \rightarrow \pi^+\pi^-$ reaction channel on $^{12}C$.  
\begin{figure}[htb]
 \centering
  \includegraphics*[angle=0,width=0.7\textwidth]{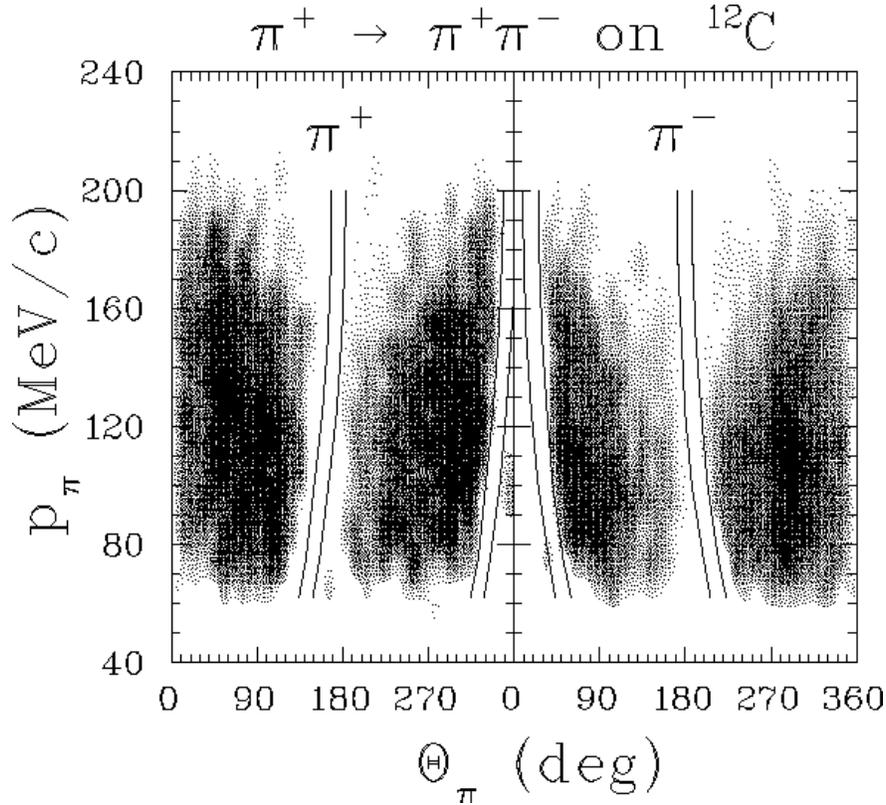}
  \setlength{\abovecaptionskip}{10pt}  
  \setlength{\belowcaptionskip}{15pt}  
  \caption{\footnotesize Diffusion plot of the CHAOS acceptance to pions 
        of the  $\pi^+ \rightarrow \pi^+\pi^-$ reaction on $^{12}C$.
        The continuous lines confine the ($p_{\pi},\Theta_{\pi}$) 
        phase space unreachable by pions, and are the result of GEANT 
        Monte Carlo simulations. The strips are 14$^\circ$ wide.}
\end{figure}
The figure also depicts the overall CHAOS acceptance for pions.  The
removal of the two CFT segments intercepting the particle beam defines 
strips in the ($p_{\pi},\Theta_{\pi}$) plane which are unreachable by 
pions. These strips are confined within bounds (continuous lines), which 
are the results of GEANT Monte Carlo simulations. The strips are 14$^\circ$
wide, 4$^\circ$ narrower than a CFT segment. The narrowing is due to the 
circular trajectories that pions acquire by moving in the magnetic field 
of CHAOS, which ultimately defines the exit direction of pions before 
impinging upon $\Delta E1$. For both positive and negative pions the 
removal of the CFT segment which allows the particle beam to enter CHAOS 
(strip at around 180$^\circ$) negligibly affects the reaction phase space. 
The missing CFT segment at the exit (strip at around 0$^\circ$) somewhat 
restricts the available $\pi \rightarrow \pi\pi$ phase space.
However, the limitation is only for an angular interval of 14$^\circ$.

The acceptance of CHAOS is irregular in the ($p_{\pi}$,$\Theta_{\pi}$)
plane due to the finite segmentation of the CFT, the two missing CFT 
segments, and the pion decays inside CHAOS. All these sources of 
non-uniformity were accounted for by assigning a {\em weight} to 
each $\pi\rightarrow\pi\pi$ event. The weights were determined using
GEANT Monte Carlo simulations, as described in more detail in 
Ref.\cite{Bonutti:three}.  Weight distributions are represented 
in Fig. 4 for the two $\pi^+\rightarrow\pi^+\pi^{\pm}$ reaction 
\begin{figure}[h]
 \centering
  \includegraphics*[angle=90,width=0.6\textwidth]{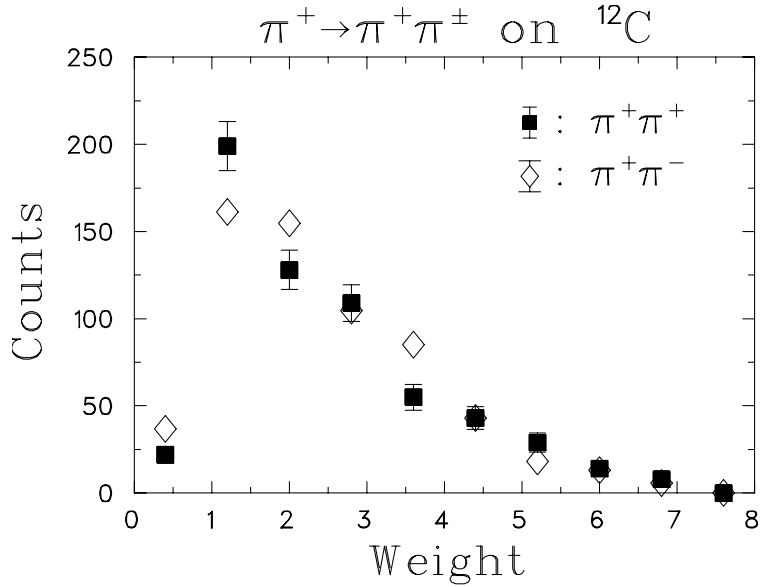}
  \setlength{\abovecaptionskip}{10pt}  
  \setlength{\belowcaptionskip}{15pt}  
  \caption{\footnotesize Weight distributions for reconstructed
   $\pi^+ \rightarrow \pi^+\pi^{\pm}$ events on $^{12}C$.}
\end{figure}
channels on $^{12}C$. These distributions show a similar 
behaviour: they are peaked around 1.5 and have a tail which extends 
up to 8. Since each $\pi\pi$ event was binned with its weight, to 
avoid large corrections in the distributions, weights were restricted 
to vary from $>$0 to $\mu+2\sigma$, where $\mu$ is the mean value 
and $\sigma$ is the standard deviation of the weight distribution. 
For $\pi^+\pi^+$ events, $\mu (\sigma)$ = 2.43(1.42), while for 
$\pi^+\pi^-$ events, $\mu (\sigma)$ = 2.52(1.31). 

\begin{figure}[h,t,b]
 \centering
  \includegraphics*[angle=90,width=0.6\textwidth]{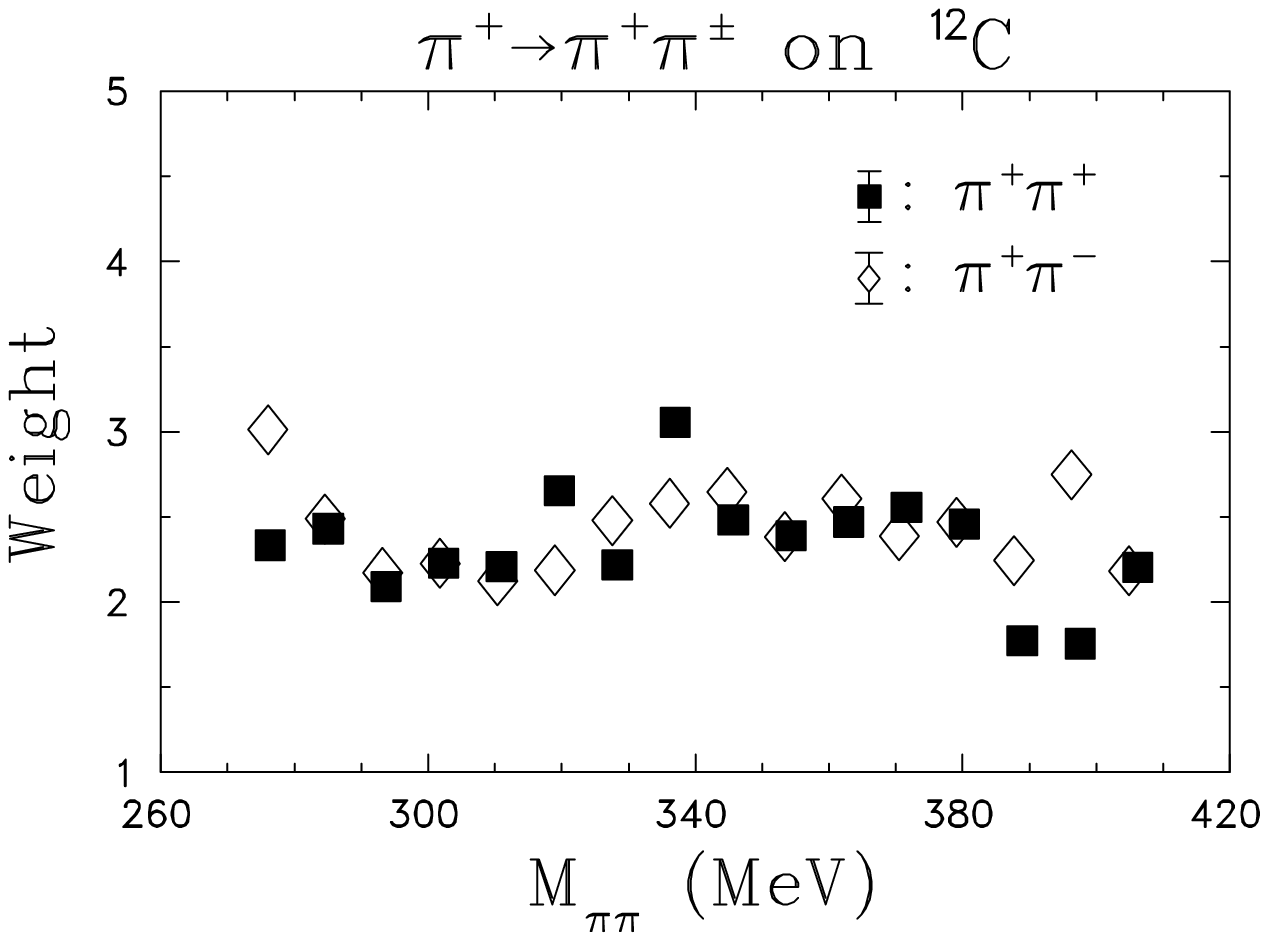}
  \setlength{\abovecaptionskip}{10pt}  
  \setlength{\belowcaptionskip}{15pt}  
  \caption{\footnotesize Weight distributions used to correct the $\pi\pi$ 
    invariant mass for the $\pi^+ \rightarrow \pi^+\pi^{\pm}$ reaction 
    channels on $^{12}C$. For a direct comparison, the $M_{\pi\pi}$ scale 
    in the present figure is the same as the scale used in Fig. 9.}
\end{figure}
As an example, in Fig. 5 are reported the weight distributions used 
to correct the measured invariant mass spectra for the two 
$\pi^+\rightarrow\pi^+\pi^{\pm}$ reaction channels on $^{12}C$. The 
average uncertainty in the weight is $\sim$6\% \cite{Bonutti:three}. 
The distributions are represented over an invariant mass interval 
where the yields of the $M_{\pi\pi}$ distributions are different from 
zero (see Fig. 9). The weight distributions appear rather similar 
and flat, especially over the shorter interval from 280 MeV to 395 
MeV, where the $M_{\pi^+\pi^{\pm}}^C$ yields are not negligible.
The flatness of the weight distributions implies that the shape of 
the measured $M_{\pi^+\pi^{\pm}}^C$ distributions changes only 
moderately when corrected for the CHAOS acceptance. The same 
conclusion applies to the other nuclei.

The angular and momentum resolutions were measured during the CHAOS 
commissioning, which just preceded the present measurement. The 
angular resolution was $<0.5^\circ$, and the momentum resolution was 1\% 
($\sigma$) for a 1 T magnetic field and 225 MeV/c pions \cite{CHAOS:one}. 
However, for the pion-production experiment, the pion mean momentum was 
$\approx$130 MeV/c (see Table 5), and the CHAOS field was set at 0.5 T. 
Simulations show that under these conditions, the angular resolution,
which was dominated by the multiple scattering, is below 2$^\circ$, 
and the momentum resolution is $\sim$4\% ($\sigma$) for 130 MeV/c pions.

2.5 Particle identification

In the case of pion-production studies, pions must be selectively 
separated from other reaction byproducts, essentially from $e, 
p$ and $d$, because the $\pi \rightarrow \pi\pi$ reaction
cross section in nuclei is from 2 to 3 orders of magnitude lower 
than the cross section of other competing reactions. For instance, 
at $T_{\pi^+} \sim$ 280 MeV and  for $^{40}Ca$, the quasi-elastic 
scattering process $\pi^+ \rightarrow \pi^+ p$ has a  total cross 
section $\sigma_{qes}\sim 800\sigma_{\pi^+\pi^+}$, while the pion 
absorption process $\pi^+ \rightarrow p p$ has a total cross section 
$\sigma_{abs} \sim 600 \sigma_{\pi^+\pi^+}$. For other processes like 
single charge exchange (scx) $\pi^+\rightarrow\pi^\circ$, the background 
arises from the $\pi^\circ \rightarrow \gamma\gamma$ decay followed by 
the $\gamma$ conversion $\gamma\rightarrow e^+e^-$ in the target or in 
the material which surrounds the target. In this case the  $e^+e^-$ 
pair may emulate a $\pi^+\pi^-$ pair.

Particles were identified by combining the information provided 
by WC's (particle momentum and polarity) and CFT's (pulse heights). 
For reactions with protons (deuterons) in the final state, i.e. 
qes and abs, the particle separation was easily achieved through 
the different response function of $\Delta E1$ and $\Delta E2$ 
to $\pi$'s and p's (d's). Pions from the pion-production process 
have a momentum distribution which does not exceed 210 MeV/c (see 
Fig. 14). At this momentum, $PH_{p}/PH_{\pi}\sim 6$, and increases 
as the momentum decreases, see Fig. 8 in Ref.\cite{CHAOS:three} 
and Fig. 3 in Ref.\cite{Bonutti:two}. However, the same method of particle
separation could not successfully be applied to $\pi$'s and e's 
for momenta exceeding 140 MeV/c.  Thus, Cerenkov counters were 
used. These counters were capable of separating $\pi$'s 
from e's with an efficiency of about 95\% for particle momenta 
up to about 230 MeV/c \cite{CHAOS:three} Fig. 11. This value increased
slightly ($\sim$98\%) when the Cerenkov pulse heights were combined 
with the $\Delta E1$ and $\Delta E2$ pulse heights. Finally, the 
$\pi$ to p and $e$ discrimination turned out to be $\sim$100\% when 
the following soft kinematic cuts were applied to particle momenta and angles: 
\begin{description}
\item[$-$] Protons which passed the pion $PH$ test were definitively 
  rejected by restricting their momenta below 210 MeV/c, the maximum 
  momentum available to pions. Such a momentum was slightly above 
  the CHAOS threshold to protons, 185 MeV/c. 
\item[$-$] Electron pairs, which might emulate $\pi^+\pi^-$ pairs, 
  appeared with opening angles $<2^\circ$. This is in accord with
  simulations of the $\pi^\circ \rightarrow \gamma\gamma$ 
  decay followed by the $\gamma$ conversion $\gamma\rightarrow e^+e^-$ 
  in CHAOS. Therefore a 2$^\circ$ cut was applied during the analysis, 
  which resulted in a $\sim$0.3\% reduction of the $\pi 2\pi$ 
  in-the-reaction plane phase space.
\end{description}
{\bf 3. Analysis} \\

The data reduction was based on fully reconstructed 
$\pi^+ \rightarrow \pi^+\pi^-$ and $\pi^+ \rightarrow \pi^+\pi^+$  
events.  Table 3 reports the yields for each channel and nucleus 
studied. In order to form differential cross sections these events 
were binned with their weights, which include the $\pi^+\pi^{\pm}$
decay rates inside CHAOS.  Some particularly useful distributions
are represented included those of the canonical observables such as 
invariant mass $M_{\pi\pi}$, angles $\Theta_{\pi}$, $\Theta_{\pi\pi}$, 
$cos\Theta_{\pi\pi}^{CM}$, and kinetic energy $T_{\pi}$. 

\begin{table}[tbc]
\caption[Table]{
Number of $\pi^{+} \rightarrow \pi^{+}\pi^{\pm}$  reconstructed events 
used  in the present analysis.
}
\begin{center}
  \begin{tabular}{lcc}                                       \hline \hline 
    \multicolumn{1}{c}{Nucleus} & 
     \multicolumn{2}{c}{Number of events [$\times 10^3$]} \\ 
           & $\pi^{+} \rightarrow \pi^{+}\pi^{-}$  
           & $\pi^{+} \rightarrow \pi^{+}\pi^{+}$         \\ \hline
$^{2}H$    &   7.27        &      1.06                    \\ 
$^{12}C$   &   6.89        &      0.62                    \\ 
$^{40}Ca$  &   4.94        &      0.49                    \\ 
$^{208}Pb$ &   3.69        &      0.31                    \\ \hline\hline
\end{tabular}
\end{center}
\end{table}

The ability to measure the kinetic energies ($T$) and laboratory 
angles ($\Theta$) of both final state pions with the CHAOS spectrometer 
permitted the determination of the five-fold differential cross section 
$\partial^5\sigma/(\partial T\partial\Theta)_{\pi_1}
(\partial T\partial\Theta)_{\pi_2}\partial\Phi_{\pi_1\pi_2}$.
Here, $\Phi_{\pi_1\pi_2}$ is the zenithal
angle between $\pi_1$ and $\pi_2$, which, in the CHAOS detector,
can be measured at either
$180^{\circ}\pm 7^{\circ}$, or $0^{\circ}\pm 7^{\circ}$.
Four-fold differential cross sections were then obtained by 
integrating out the $\Phi_{\pi\pi}$ dependence. This was accomplished
by using a linear function joining the two measured data points,
and the result was the factor $f_{\Phi}$. This method of assessing 
$f_{\Phi}$ leads to a systematic uncertainty of 8\% 
($\sigma$) for deuterium \cite{Bonutti:three} and 6-7\% 
($\sigma$) for nuclei. The data were also reduced to 
single differential cross sections $d\sigma/d{\cal O}_{\pi}$, 
where ${\cal O}_{\pi}$ represents $(T_{\pi}$ or $\Theta_{\pi})_{1,2}$. 
The cross section was then related 
to measured quantities $\frac{d\sigma}{d{\cal O}_{\pi}}$=
$f_e\frac{N({\cal O}_\pi)}{\Delta{\cal O}_\pi}$, where $f_e$ is 
a parameter which is determined by the experimental conditions, 
$N({\cal O}_\pi)$ is the number of weighted events in a given bin 
and $\Delta{\cal O}_\pi$ is the bin width for the ${\cal O}_\pi$ 
observable. The experimental parameter $f_e$ is the product of 
several factors which were individually measured in order to 
obtain absolute values for the cross sections: $f_e = f_{\Phi}
(N_{\pi} {\cal C}_l \mathcal{E}_{WC} N_{sc})^{-1}$, where 
$N_{\pi}$ is the number of pions impinging upon the target (see 
discussion in Paragraph 2.1), ${\cal C}_l$ is the computer live-time 
during data acquisition which was 61\% for $^{2}H$, 53\% for $^{12}C$,
70\% for $^{40}Ca$ and 80\% for $^{208}Pb$, $\mathcal{E}_{WC}$ is the 
overall WC efficiency 49.5\%, and $N_{sc}$ is the number of target 
scattering centres per unit area (see discussion in Paragraph 2.2). 
The uncertainty in $f_e$ $\sim$11\% ($\sigma$) comes mostly from the 
uncertainties in $\mathcal{E}_{WC}$ and $f_{\Phi}$.  In fact, 
$\mathcal{E}_{WC}=
\mathcal{E}_{WC1}\mathcal{E}_{WC2}(\mathcal{E}_{WC1}
\mathcal{E}_{WC2}\mathcal{E}_{WC3})^2$, and the average uncertainty 
in measuring the efficiency of a wire chamber was 0.93\% thus yielding
an overall $\mathcal{E}_{WC}$ uncertainty of 8.3\%. The WC4 
efficiency uncertainty was 0.0\% since only three out of eight 
independent anodes of one cell were required to fire in order to 
determine the direction of an outgoing pion. Finally, the overall 
uncertainty was obtained by summing the uncertainty due to $f_{\Phi}$ 
and $\mathcal{E}_{WC}$ in quadrature.

{\bf 4. Models of the $\pi 2\pi$ reaction and the $\pi\pi$ 
        interaction in the nuclear medium} \\
 
This experiment employed the $\pi^+ A\rightarrow\pi^+\pi^{\pm}A'$ 
reaction ($\pi 2\pi$) to study the near-threshold $\pi\pi$ 
interaction as a function of the (average) nuclear density. Other
reactions can, in principle, be used for similar studies, eg. the 
$p \rightarrow \pi\pi$ and $\gamma \rightarrow \pi\pi$ reactions.
For any reaction and  for a reliable interpretation of the $\pi\pi$ 
dynamics in nuclear matter, theoretical models should account for 
both the reaction mechanism, and nuclear distortions (i.e. pion 
absorption and others). In the case of $\pi 2\pi$, this approach 
was followed in the work of Refs. \cite{Vicente:one,Rapp:one},  
which also included the kinematical limits of the experimental 
apparatus. For those theoretical works which
addressed only the dynamics of a $(\pi\pi)_{I=J=0}$ interacting system 
in nuclear matter, it was useful to define the observable
$\cal C$$_{\pi\pi}^A$ (see Paragraph 5.5) since it is only
weakly dependent on the reaction mechanism and 
nuclear distortions \cite{Bonutti:four}. 

Before discussing the data and the comparison with the available 
theoretical calculations for both the $\pi 2\pi$ reaction and the 
behaviour of a $\pi\pi$ interacting system in nuclear matter, some 
details of the models are presented below.

\begin{description}
\begin{figure}[htb]
 \centering
  \includegraphics*[angle=0,width=0.8\textwidth]{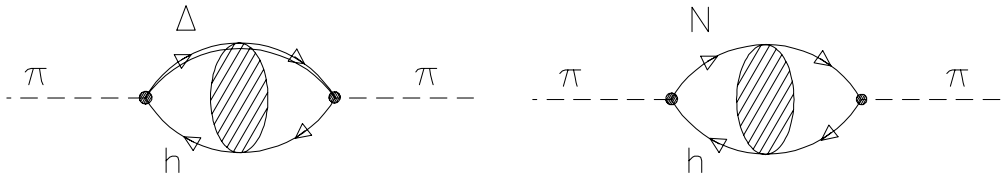}
  \setlength{\abovecaptionskip}{10pt}  
  \setlength{\belowcaptionskip}{15pt}  
  \caption{\footnotesize Diagrams depicting the P-wave coupling of pions
    to $p-h$ and $\Delta-h$ correlated states.}
\end{figure}
 \item[$\mathcal{M} 1$:] In order to understand the main features
    of medium modification on the $\pi\pi$ interacting system
    \cite{Rapp:one}, the pion-production reaction on nuclei is modelled 
    by means of the two leading reactions: 
    1) the one-pion exchange (OPE) reaction,
    $\pi^+ N \rightarrow \pi^+ \pi^{\pm} N$, which contributes to 
    both the isoscalar and isotensor channels, and 
    2) the $N^*(1440)$ resonance excitation which is restricted to 
    decay only to isoscalar $\pi\pi$ states, 
    $\pi^+ N\rightarrow N^*\rightarrow\epsilon N\rightarrow N(\pi\pi)_{I=J=0}$.
     The in-vacuum $\pi\pi$ interaction is accounted for by the 
    chirally-improved J\"{u}lich model\cite{Julich:one}, which in 
    the GeV region is able to describe the $\pi\pi\rightarrow\pi\pi$ 
    scattering and to reproduce the $I=J=0$ $\pi\pi$ phase shifts. 
    The $S-$wave $\pi\pi$ (final state) interaction is modified in 
    the nuclear medium through standard renormalization of the pion 
    progagator (see Fig. 6) which determines the following: in the 
    $\pi^+\pi^-$ channel, the isoscalar $\pi\pi$ amplitude is strongly 
    reshaped which finally provides the near-threshold $M_{\pi^+\pi^-}$ 
    enhancement observed in the CHAOS data; in the $\pi^+\pi^+$ channel, 
    the nuclear density has little effect on the pure isotensor $\pi\pi$ 
    amplitude. In order to have a realistic comparison with the CHAOS 
    results, the model deals with common nuclear effects like Fermi 
    motion and pion absorption although using approximate approaches, 
    and calculates the many-fold differential cross sections by accounting 
    for the CHAOS acceptance.
  \item[$\mathcal{M} 2$:] The effort in Ref.\cite{Vicente:one} is 
    to model the $\pi^+ A \rightarrow \pi^+ \pi^{\pm} A'$ reaction
    through a microscopic description of the elementary
    $\pi N\rightarrow \pi\pi N$ reaction and a detailed study of
    the production reaction in nuclei.
\begin{figure}[htb]
 \centering
  \includegraphics*[angle=90,width=0.6\textwidth]{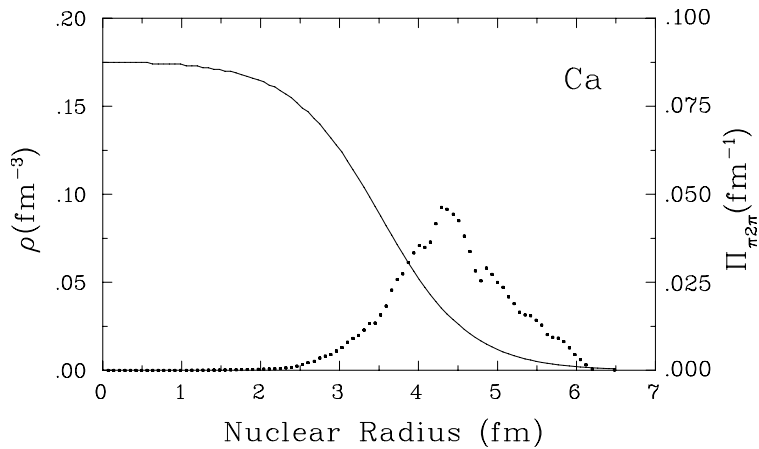}
  \setlength{\abovecaptionskip}{10pt}  
  \setlength{\belowcaptionskip}{15pt}  
  \caption{\footnotesize Nuclear density distribution ($\rho$, solid curve) 
      and probability of a $\pi 2 \pi$ event to occur ($\Pi_{\pi 2\pi}$, 
      dotted curve) for $Ca$ as a function of the nuclear radius. The area
      subtended by $\Pi_{\pi 2\pi}$ is normalized to 1.}
\end{figure}
    The elementary pion-production reaction relies on five Feynmann 
    diagrams, which have $N$'s, $\Delta$'s and $N^*$'s as intermediate 
    isobars. The relative scattering amplitudes are derived from chiral 
    lagrangian. The purely mesonic $\pi\pi$ amplitude is calculated by 
    means of the coupled-channel Bethe-Salpeter equation, and the 
    resulting amplitude is capable of predicting the experimental phase 
    shifts of the scalar-isoscalar channel up to 1.2 GeV. With this 
    detailed approach to the $\pi N\rightarrow \pi\pi N$ reaction, the 
    model is able to predict the total cross sections for the $\pi^+\pi^-$, 
    $\pi^{\circ}\pi^{\circ}$ and $\pi^+\pi^+$ channels from threshold up 
    to 300 MeV. In the nuclear medium, the $(\pi\pi)_{I=J=0}$ interaction 
    is strongly modified due to the coupling of pions to $p-h$ and 
    $\Delta-h$ excitations, which are represented by the diagrams in 
    Fig. 6. This results in a displacement of the $ImT_{\pi\pi}$ strength 
    toward the $2m_{\pi}$ threshold as the nuclear density increases. 
    Conversely, the nuclear medium weakly affects the 
    $(\pi\pi)_{I=2,J=0}$ interaction.

    The model includes several nuclear effects: Fermi motion, Pauli 
    blocking, pion absorption and quasi-elastic scattering. The last two 
    interactions are found to largely remove $\pi 2\pi$ pions from 
    the incident flux, which ultimately confines the pion-production 
    process on the nuclear skin\cite{Vicente:two}. Fig. 7 shows the 
    probability of a $\pi 2\pi$ event to occur ($\Pi_{\pi 2\pi}$, dotted 
    curve) along with the nuclear density distribution ($\rho$, solid 
    curve) for $Ca$ as a function of the nuclear radius. The two 
    distributions yield a weighted mean nuclear density of $\rho$
    =0.24$\rho_n$, which qualitatively agrees with the experimental
    result reported in\cite{Bonutti:two}, where the interacting 
    nucleon of the $\pi N[A-1]\rightarrow \pi\pi N[A'-1]$ reaction is 
    found to preferentially lie in an external nuclear orbit. The above 
    value of $\rho$ can also be used for comparison with the nuclear 
    densities quoted in Ref.\cite{Rapp:one} and discussed in the 
    Paragraph 5.1. Finally, the model employs a Monte Carlo
    technique to generate and propagate $\pi 2\pi$ events, thus their 
    phase space can easily accommodate the CHAOS acceptance.
  \item[$\mathcal{M} 3$:] The work of Ref.\cite{Hatsuda:one} aims at
    studing the possibility of the $\sigma-$meson identification in 
    nuclear matter, where the elusive particle might be detected as a
    consequence of the partial restoration of chiral symmetry. Due 
    to the strong interaction of $\sigma$ with the nuclear medium,
    the description of the $\sigma$ properties in terms of parameters 
    like $m_{\sigma}$ and $\Gamma_{\sigma}$ appears inadequate, and a 
    proper observable becomes the $\sigma$ spectral function $\rho_\sigma$. 
    By using a simple but general approach, \cite{Hatsuda:one} proves 
    that in the proximity of the $2m_\pi$ threshold the partial 
    restoration of the chiral symmetry implies
    $\rho_\sigma(\omega\sim 2m_\pi)=-\pi^{-1}/Im\Sigma_\sigma(\omega,\rho)$, 
    where $\Sigma_\sigma$ is the $\sigma$ self-energy, $\omega$ is
    the total energy and $\rho$ the nuclear density. Near $2m_\pi$ 
    threshold, the $Im\Sigma_\sigma$ is proportional to the phase 
    space factor $(1 - \frac{2m_\pi^2}{\omega^2})^{1/2}$, thus 
    $\rho_\sigma (\omega\sim 2m_\pi) \propto \theta(\omega - 2m_\pi) / 
    (1 - \frac{2m_\pi^2}{\omega^2})^{1/2}$. A direct consequence is that 
    the $\sigma$ spectral function strength enhances as the $\sigma$ total 
    energy approaches $2m_\pi$. In order to render this general finding 
    more quantitative, the theory uses the SU(2) linear sigma model. The 
    mean field correction in nuclear matter of $\Sigma_\sigma(\rho)$ 
    is accounted for by the leading diagram sketched in Fig. 8. 
    Furthermore, the model does not include collective pionic modes to 
    derive a possible source of near-threshold strength from the pion 
    coupling to $p-h$ and $\Delta-h$ correlated states. 
  \item[$\mathcal{M} 4$:] This work\cite{Aouissat:two} studies the 
    $(\pi\pi)_{I=J=0}$ interaction in nuclear matter by constructing 
    a microscopic theory for both  the $\sigma-$meson propagator
    ($D_{\sigma}$) and the $\pi\pi$ $T-$matrices in the framework of 
    the linear sigma model. In this approach the basic chiral symmetry 
    constraints, i.e. the vanishing of the scattering length in the 
    chiral limit and the Ward's identities are satisfied, and in addition, 
    the theory is capable of reproducing the experimental phase shift 
    in the scalar-isoscalar channel. In-medium pions are renormalized 
    through their $P-$wave coupling to correlated $p-h$ and $\Delta-h$ 
    states, as depicted in Fig. 6. Bare sigmas, of the linear sigma model, 
    are coupled to the nuclear medium via the tad-pole diagram, which 
    is sketched in Fig. 8. The $P-$wave renormalization of sigmas leads 
    to a downward shifts of the $\sigma$ mass, which ultimately produces 
    a strong enhancement of the $\sigma-$meson mass distribution (i.e. 
    $ImD_{\sigma}$) around the $2m_{\pi}$ threshold. This enhancement 
    is further increased by a factor of 2 to 3 by the $S-$wave 
    renormalization of the bare $\sigma-$meson mass (i.e. by the partial 
    restoration of chiral symmetry), which presents a similar behavior 
    near the $2m_{\pi}$ threshold.
\begin{figure}[htb]
 \centering
  \includegraphics*[angle=0,width=0.2\textwidth]{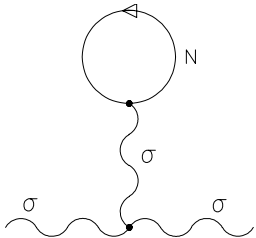}
  \setlength{\abovecaptionskip}{10pt}  
  \setlength{\belowcaptionskip}{15pt}  
  \caption{\footnotesize Diagram (tad-pole) describing the $S-$wave coupling
           of $\sigma$ to the nuclear medium.}
\end{figure}
\end{description}

{\bf 5. Results of the $\pi \rightarrow \pi\pi$ reaction in nuclei} \\

A general property of the $\pi 2\pi$ process on nuclei in the low-energy
$M_{\pi\pi}$ regime was outlined by previous experimental works: it is
a quasi-free process both when it occurs on deuterium \cite{Rui:one} and 
on complex nuclei \cite{Bonutti:two}. Furthermore, a common reaction 
mechanism underlies the process whether it occurs on a nucleon or a 
nucleus \cite{Bonutti:four}. Thus the study of the 
$\pi^+$ $^2H \rightarrow \pi^+\pi^{\pm} NN$ reaction is dynamically 
equivalent to studying the elementary $\pi^+ n \rightarrow \pi^+\pi^- p$ 
and $\pi^+ p \rightarrow \pi^+\pi^+ n$ reactions separately.

\begin{figure}[t]
 \centering
  \includegraphics*[angle=0,width=0.5\textwidth]{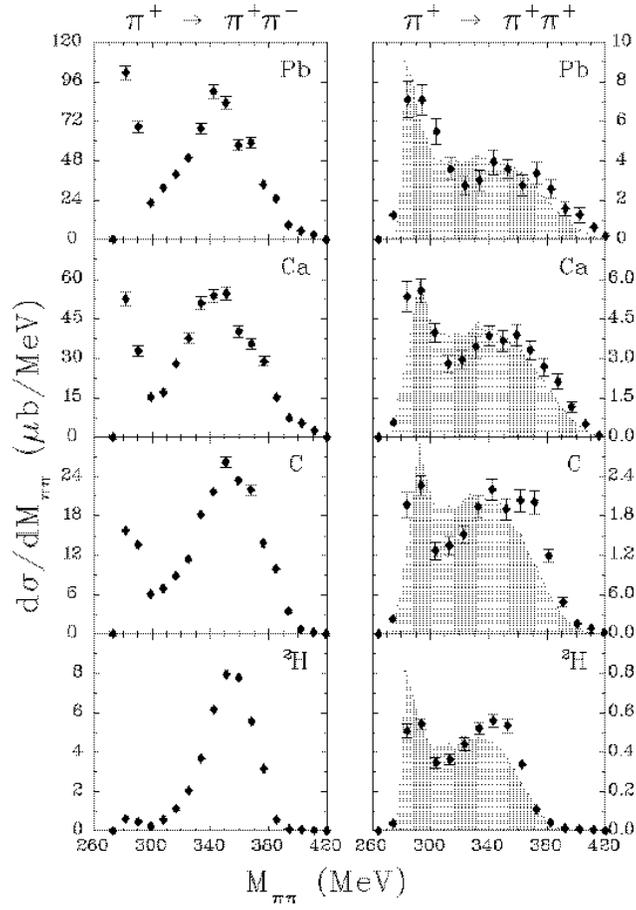}
  \setlength{\abovecaptionskip}{10pt}  
  \setlength{\belowcaptionskip}{15pt}  
  \caption{\footnotesize Invariant mass distributions (diamonds) for 
      the $\pi^+ \rightarrow \pi^+\pi^-$ and $\pi^+ \rightarrow \pi^+\pi^+$ 
      reactions on $^{2}H$, $^{12}C$, $^{40}Ca$ and $^{208}Pb$.
      The shaded regions represent the results of phase-space simulations
      for the pion-production reaction $\pi A \rightarrow \pi\pi N [A-1]$.} 
\end{figure}

In the present measurement, the $\pi^+ A\rightarrow \pi^+\pi^{\pm} A'$ 
reactions were studied under the same experimental conditions. Thus
for a given observable the distributions are directly comparable. The 
error bars explicitly shown on the spectra represent the statistical 
uncertainties, which must be added in quadrature to the systematic one 
$\sim$11\% ($\sigma$) to obtain the overall uncertainty associated with 
the data points. The only exception is for the $cos\Theta_{\pi\pi}$ 
distributions in Fig. 10, where the overall uncertainties are shown.
\begin{figure}[t]
 \centering
  \includegraphics*[angle=0,width=0.5\textwidth]{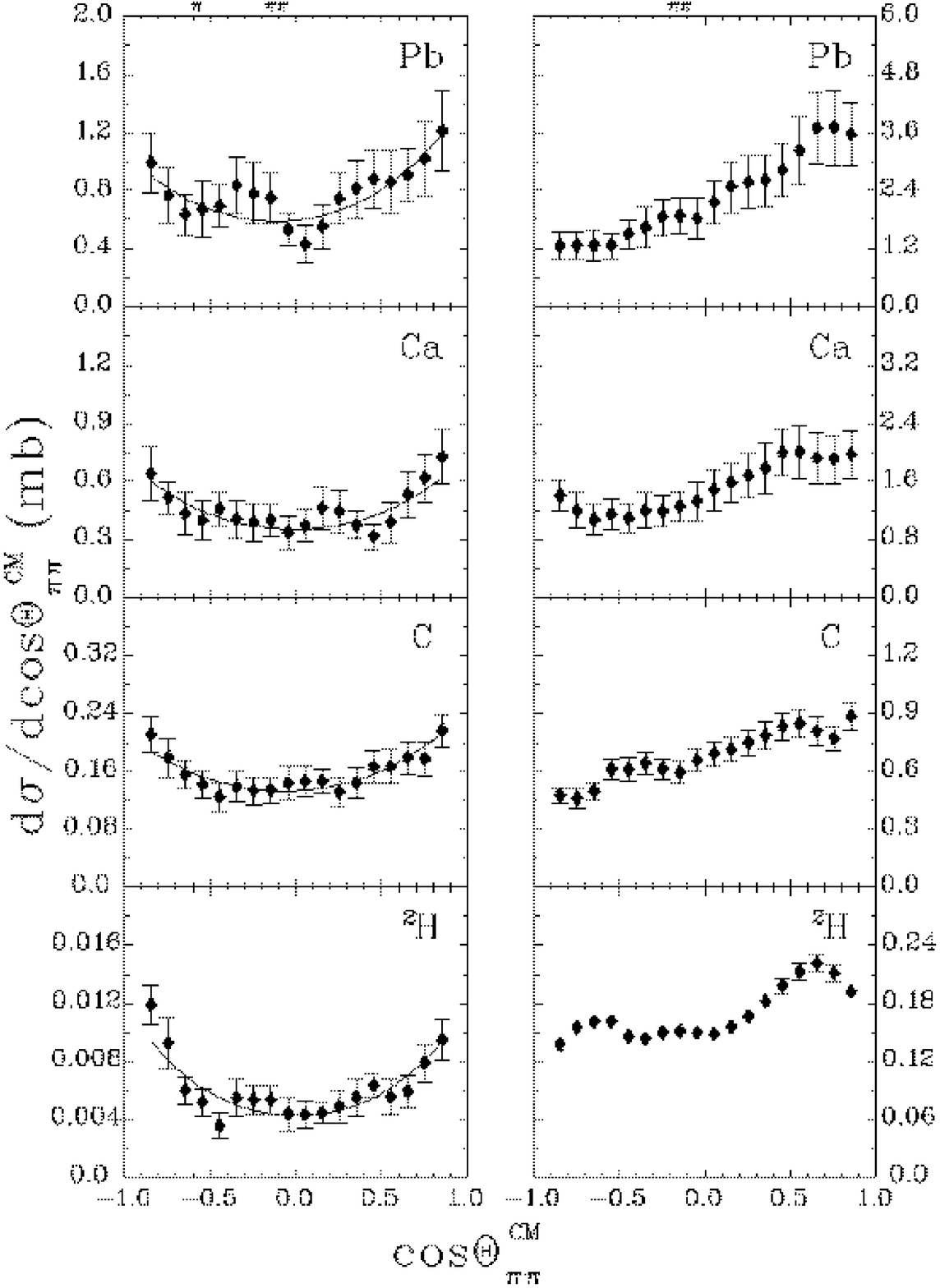}
  \setlength{\abovecaptionskip}{10pt}  
  \setlength{\belowcaptionskip}{15pt}  
  \caption{\footnotesize Distribution of cos$\Theta$ 
    in the $\pi^+\pi^-$ centre-of-mass frame for the 
    $\pi^+ A \rightarrow \pi^+\pi^- A'$ reaction for
    $2m_{\pi}\leq M_{\pi\pi}\leq$ 310 MeV (left frame), and 
    for $310 < M_{\pi\pi}\leq$ 420 MeV (right frame). The
    solid lines are best-fits to the data including S, 
    P and D waves. }
\end{figure}

5.1 The $\pi\pi$ invariant mass 

Fig. 9 shows the single differential cross sections (diamonds) as 
a function of the $\pi\pi$ invariant mass ($M_{\pi\pi}$, MeV) for the 
two reaction channels
$\pi^+ \rightarrow \pi^+\pi^-$ and $\pi^+ \rightarrow \pi^+\pi^+$ .
Horizontal error bars are not indicated since 
they lie within symbols. The distributions span the total energy interval 
available to the $\pi 2\pi$ reaction, which ranges from $2m_{\pi}$, 
the low-energy threshold, up the 420 MeV, the maximum allowed by the 
reaction. The $\pi A \rightarrow \pi\pi N [A-1]$ phase space simulations 
(dotted histograms for $A:$ $^{2}H$, $^{12}C$, $^{40}Ca$ and $^{208}Pb$) 
are also shown, and are normalized to the same area as the 
experimental distributions. 

Regardless of the nuclear mass number, the invariant mass distributions for
the $\pi^+ \rightarrow \pi^+\pi^+$ channel closely follow phase 
space, and the energy maximum increases with increasing $A$, 
that is, with increasing nuclear Fermi momentum. The 
$\pi^+\rightarrow\pi^+\pi^-$ channel displays a different behaviour.
Compared to phase space, the $^{2}H$ invariant mass displays little 
strength from $2m_{\pi}$ to 310 MeV, while, in the same energy interval, 
the $^{12}C$, $^{40}Ca$ and $^{208}Pb$ $\pi^+\pi^-$ invariant mass 
distributions increasingly peak as $A$ increases. 

In order to explain the nature of the reaction mechanism 
contributing to the peak structure, it is useful to examine
cos$\Theta_{\pi\pi}^{CM}$ distributions and $(p_{\pi^+},p_{\pi^-})$ 
diffusion plots for those events with
invariant masses in the region of the peak.
$\Theta_{\pi\pi}^{CM}$ is the angle between the direction of a final pion 
and the direction of the incoming pion beam in the $\pi^+\pi^-$ rest frame. 
Fig. 10 shows the cos$\Theta_{\pi\pi}^{CM}$ distributions (diamonds) for 
$2m_{\pi}\leq M_{\pi^+\pi^-}\leq 310$ MeV and $310 < M_{\pi^+\pi^-}\leq 420$ 
MeV, the latter being shown for comparison.
The vertical error bars represent the overall uncertainties, which are 
the systematic and statistical uncertainties summed in quadrature. The
solid lines in Fig. 10 represent the best-fit to the differential
distributions, which was obtained with a partial 
wave expansion limited to the three lowest waves, i.e. S, P and D. The 
best-fit results are reported in Table 4, along with $\chi^2_{\nu}$,
which was evaluated using the overall uncertainties. For all the nuclei
studied $\chi^2_{\nu}\leq 1$ which indicates that a proper number of waves 
was used in the expansion. In the case of heavier nuclei, the $\pi^+\pi^-$ 
system predominantly couples in $S-$wave $\sim$ 95\% (or $\ell$=0
relative angular 
momentum) and a remaining 5\% is spent in a $D-$wave ($\ell$=2) state. 
Furthermore, within the sensitivity of the $\chi^2_{\nu}-$method, any 
$P-$wave ($\ell$=1) coupling of the two pions is excluded. 
\begin{table}[htb]
\caption[Table]{$S-$ $P-$ and $D-$wave contributions to the 
  $\pi^{+} \rightarrow \pi^{+}\pi^{-}$ reaction channel 
  for $2m_{\pi} \leq M_{\pi^+\pi^-} \leq$ 310 MeV.}

\begin{center}
  \begin{tabular}{lccccccc}   \hline \hline 
Nucleus    & $S-$wave (\%) & &$P-$wave (\%) & &$D-$wave (\%) & & $\chi^2_{\nu}$ \\ \hline 

$^{2}H$    &   88.2      & &   0.0        & &    11.8      & &    1.03        \\
$^{12}C$   &   96.0      & &   0.0        & &     4.0      & &    0.29        \\
$^{40}Ca$  &   93.9      & &   0.0        & &     6.1      & &    0.43        \\
$^{208}Pb$ &   93.9      & &   0.8        & &     5.3      & &    0.46        \\ 
\hline\hline
\end{tabular}
\end{center}
\end{table}
\begin{figure}[t]
 \centering
  \includegraphics*[angle=0,width=0.5\textwidth]{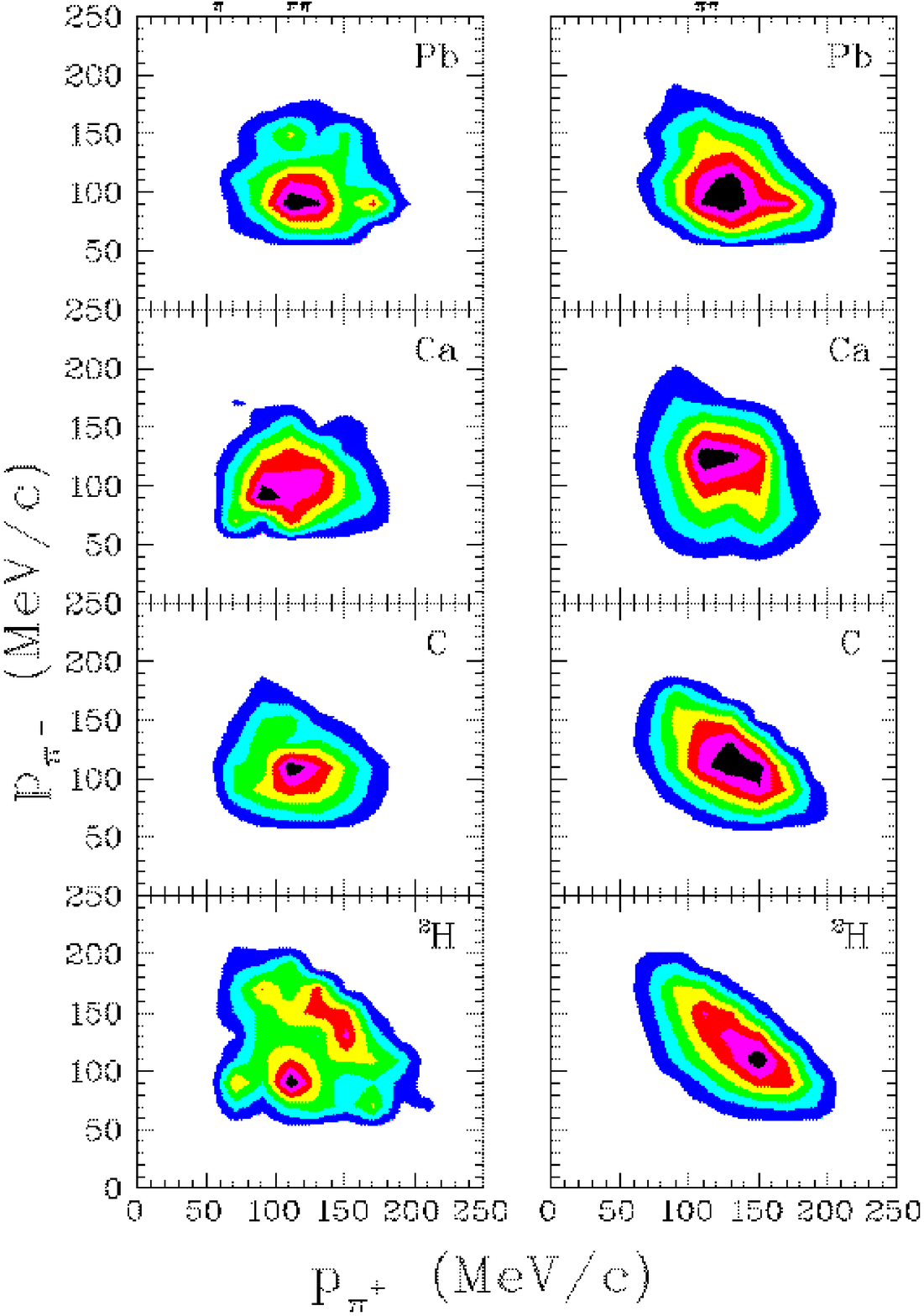}
  \setlength{\abovecaptionskip}{10pt}  
  \setlength{\belowcaptionskip}{15pt}  
  \caption{\footnotesize Diffusion plots of final pion momenta for
      the $\pi^+ \rightarrow \pi^+\pi^-$ reaction channel. The
      $\pi^+\pi^-$ invariant mass is limited to the peak region 
      $2m_{\pi}\leq M_{\pi\pi}\leq$ 310 MeV (left frame), and to the
      region above it 310$< M_{\pi\pi}\leq$ 420 MeV (right frame).}
\end{figure}
Fig. 11 displays the diffusion plots of final pion momenta for the 
$\pi^+ \rightarrow \pi^+\pi^-$ reaction channel. For the plots on the 
left side, $M_{\pi^+\pi^-}$ is constrained to the peak region, i.e.
$2m_{\pi}\leq M_{\pi\pi}\leq$ 310 MeV. The plots on right side,
reported for comparison, are for invariant masses in the higher
energy region, 310$<M_{\pi\pi}\leq$420 MeV. In the peak 
region, $^2H$ has the pion momenta flatly diffused over the available 
phase space, while for the more complex nuclei, the choice of
$M_{\pi^+\pi^-}\le$310 MeV highlights a bump structure in the momentum 
range 90-140 MeV/c. Generally, the ($p_{\pi^+},p_{\pi^-}$) 
diffusion plots on the right side have a peak-structure which is somewhat 
wider and centered at higher pion momenta.
\begin{figure}[t]
 \centering
  \includegraphics*[angle=270,width=0.6\textwidth]{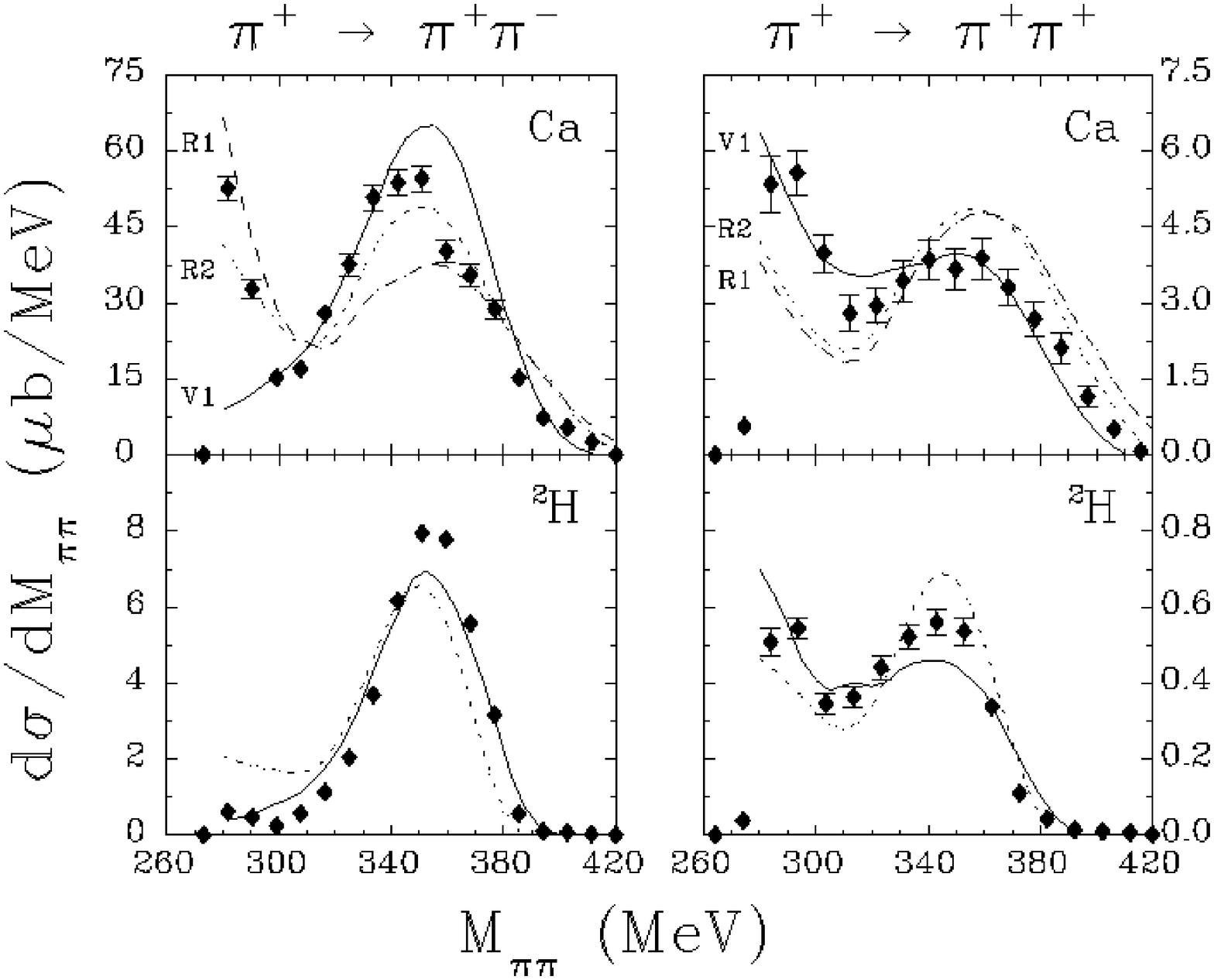}
  \setlength{\abovecaptionskip}{10pt}  
  \setlength{\belowcaptionskip}{15pt}  
  \caption{\footnotesize Invariant mass distributions (diamonds) for 
      the $\pi^+ \rightarrow \pi^+\pi^-$ and $\pi^+ \rightarrow \pi^+\pi^+$ 
      reactions on $^{2}H$ and $^{40}Ca$. The dashed curves are taken from 
      Ref.\cite{Rapp:one} and the model is explained in $\mathcal{M} 1$, 
      $R1$ and $R2$ are the results for 0.7$\rho_n$
      and 0.5$\rho_n$, respectively. The full curves ($V1$) are from the
      theoretical work of Ref.\cite{Vicente:one} $\mathcal{M} 2 $ with
      a mean $\rho$=0.24$\rho_n$. The curves are normalized to the 
      experimental data.}
\end{figure}

The results of
two recent theoretical works, $\mathcal{M}1$ \cite{Rapp:one} and 
$\mathcal{M} 2$ \cite{Vicente:one}, which have modelled the $\pi 2 \pi$ 
reaction on nuclei, are reported in Fig. 12. The (short 
and long) dashed lines denote the $\mathcal{M}1$ calculations while 
the $\mathcal{M}2$ predictions are shown as full 
lines. In the case of $Ca$, $R1$ ($R2$) indicates the $\mathcal{M}1$ 
predictions for $\rho$=0.7$\rho_n$ ($\rho$=0.5$\rho_n$), where 
$\rho_n$ is the nuclear saturation density, while $V1$ is the result 
of the $\mathcal{M}2$ calculations for a mean $\rho$=0.24$\rho_n$.
For the purpose of the present discussion, the curves in Fig. 12 
are normalized to the experimental data. 

For the $\pi^+ \rightarrow \pi^+ \pi^+$ channel, the $R1$ and $R2$ 
distributions differ slightly from each other.  The $R1$ distribution is 
broader, due to the larger nuclear Fermi momentum, which is a consequence 
of the higher nuclear density used \cite{Rapp:one}. The lower $\rho$ used 
for $R1$ seems to describe the measured distribution better. This trend is 
also supported by $V1$ whose $\mathcal{M} 2$ predictions agree well with 
the invariant mass distributions. Therefore, both $\mathcal{M}1$ and 
$\mathcal{M}2$ indicate that the average nuclear density of $^{40}Ca$ for 
which the production reaction takes place cannot exceed 0.5$\rho_n$.

In the $\pi^+ \rightarrow \pi^+ \pi^-$ channel the distribution 
predicted by $\mathcal{M}2$ for $^{2}H$ is able to describe the data, 
although it tends to overestimate the low-energy $M_{\pi\pi}$ yield. The 
present version of the model is considerably improved compared to previous 
versions \cite{Oset:one,Bonutti:three}, thus placing the construction
of nuclear medium effects on more reliable grounds. The $\mathcal{M}2$ 
approach to $^{40}Ca$ includes several medium effects: Fermi motion, 
pion absorption, pion quasi-elastic scattering, and $(\pi\pi)_{I=J=0}$
medium modifications.  Nevertheless, the model is unable to reproduce
the observed $M_{\pi\pi}$ strength in the near threshold region. An
increase of $\rho$ from 0.24$\rho_n$ to 0.5$\rho_n$ and to 0.7$\rho_n$ 
is unlikely to improve the agreement with the experimental cross section,
see also Fig. 9 of Ref.\cite{Vicente:one}. In the case of $\mathcal{M}1$, 
the $R2$ prediction (0.5$\rho_n$) seems to reproduce the $M_{\pi\pi}$ 
distribution better, although some of the near-threshold yield already 
comes from the $\pi^+$ $^2H \rightarrow \pi^+ \pi^- pp$ production. The 
model, in fact, overestimates the cross section at low invariant masses 
whose amount is shown in Fig. 12. The $R1$ solution (0.7$\rho_n$) provides 
the correct $M_{\pi\pi}$ near-threshold intensity but does not describe 
the remaining part of the distribution very well. Therefore, $V1$, $R1$ 
and $R2$ suggest that the missing $M_{\pi^+\pi^-}$ strength near the 
2m$_\pi$ threshold should be searched for in a stronger $\rho$-dependence 
of the $(\pi\pi)_{I=J=0}$ interaction, rather than by requiring the 
pion-production process to occur in an unlikely high-density nuclear 
environment.

\begin{figure}[t]
 \centering
  \includegraphics*[angle=0,width=0.5\textwidth]{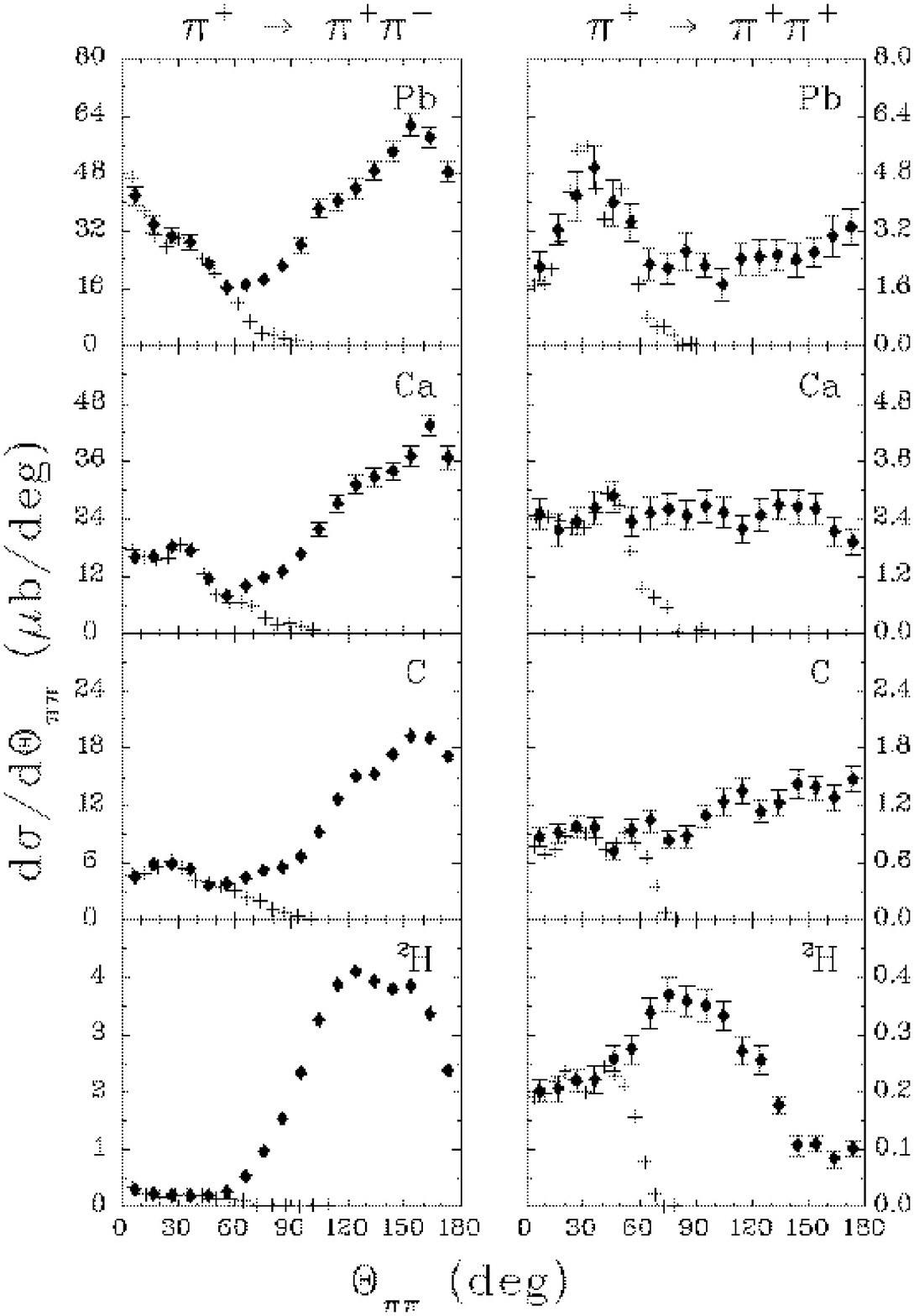}
  \setlength{\abovecaptionskip}{10pt}  
  \setlength{\belowcaptionskip}{15pt}  
  \caption{\footnotesize Opening angle distributions for the 
      $\pi^+ \rightarrow \pi^+\pi^-$ and $\pi^+ \rightarrow \pi^+\pi^+$ 
      reactions on $^{2}H$, $^{12}C$, $^{40}Ca$ and $^{208}Pb$ for 
      $2m_{\pi}\leq M_{\pi\pi}\leq$ 310 MeV (crosses) and for 
      $2m_{\pi}\leq M_{\pi\pi}\leq$ 420 MeV (diamonds).}
\end{figure}

5.2 The $\pi\pi$ opening angle

For the $\pi A \rightarrow \pi_1 \pi_2 A'$ production reaction, the
relation between the $\pi_1 \pi_2$ invariant mass and the directly
measured quantities is:
$M^{2}_{\pi_1 \pi_2}$ =  
4$m^{2}_{\pi} + 2T_{\pi_1}T_{\pi_2} + 2m_{\pi}(T_{\pi_1}+T_{\pi_2}) 
- p_{\pi_1}p_{\pi_2}cos \Theta_{\pi_1 \pi_2}$, where $T_{\pi} (p_{\pi})$ 
is the pion kinetic energy (momentum) and $\Theta_{\pi_1 \pi_2}$ is 
the opening angle of the pion pair. The behaviour of $\Theta_{\pi_1\pi_2}$ 
as a function of $M_{\pi_1\pi_2}$ is studied by directly applying 
soft-cuts on $M_{\pi_1 \pi_2}$. Fig. 13 illustrates the differential 
cross section as a function of $\Theta_{\pi\pi}$ for 
$2m_\pi\leq M_{\pi\pi}\leq$310 MeV (crosses), and 
$2m_{\pi}\leq M_{\pi\pi}\leq$ 420 MeV (diamonds). The 
differential cross sections nearly coincide for all nuclei from 
0$^\circ$ up to 60$^\circ$, which indicates that the near-threshold 
pion pairs are preferentially emitted at small opening angles, regardless 
of both the pion energy and the reaction channel. In general, the 
$\Theta_{\pi\pi}$ distributions for the two reaction channels vary 
markedly with $A$, making this observable suitable for selective 
tests of various $\pi 2\pi$ models.

5.3 The $\pi$ energy

\begin{figure}[htb]
 \centering
  \includegraphics*[angle=0,width=0.5\textwidth]{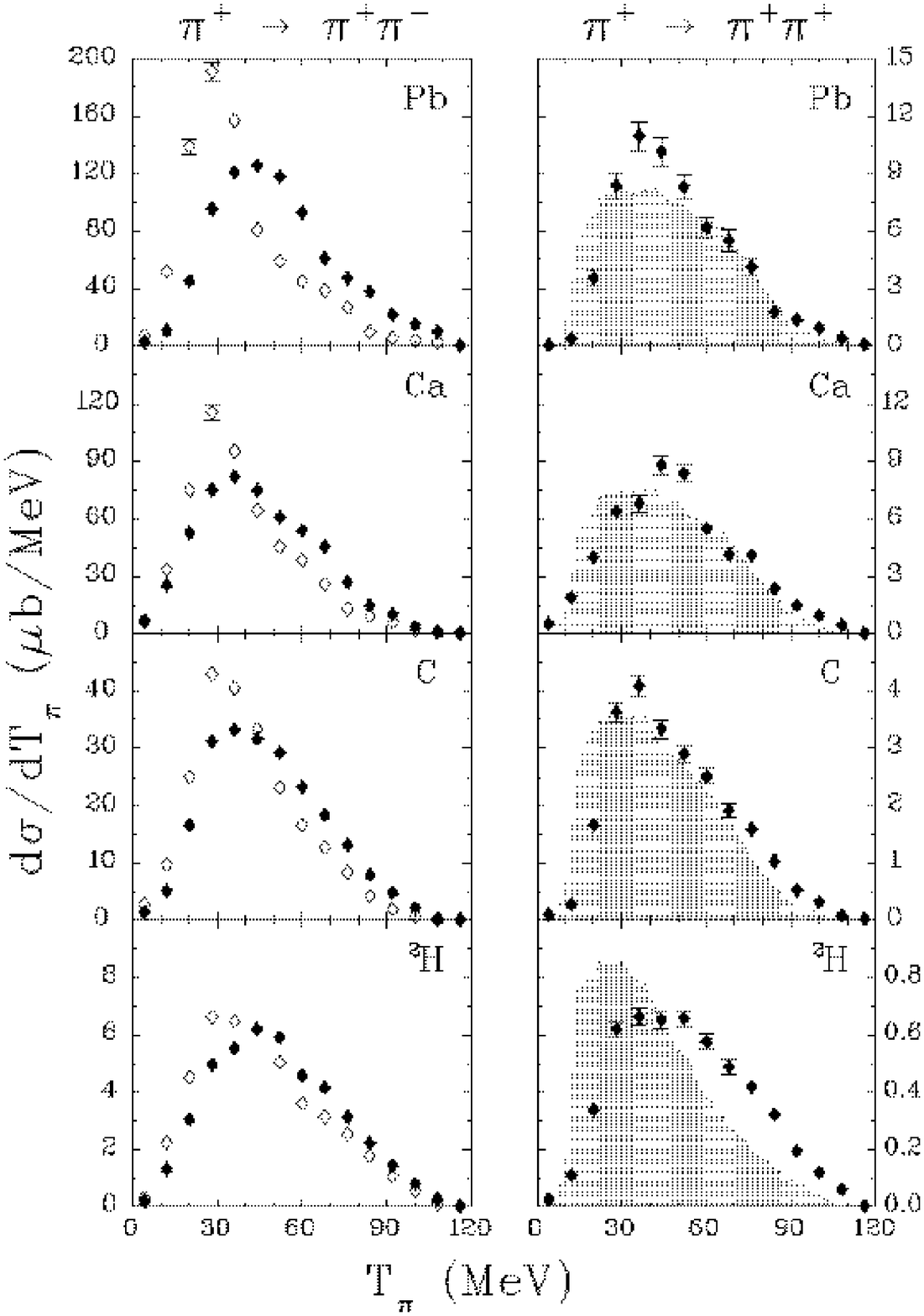}
  \setlength{\abovecaptionskip}{10pt}  
  \setlength{\belowcaptionskip}{15pt}  
  \caption{\footnotesize Single pion kinetic energy distributions for the 
      $\pi^+ \rightarrow \pi^+\pi^-$ and $\pi^+ \rightarrow \pi^+\pi^+$ 
      reactions on $^{2}H$, $^{12}C$, $^{40}Ca$ and $^{208}Pb$. Positive 
      and negative pion distributions are denoted with full and open
      diamonds, respectively. The shaded areas represent the results of 
      $\pi^{+} A \rightarrow \pi^{+}\pi^{+} n [A-1]$ simulations. The 
      CHAOS threshold for pions is 11 MeV.}
\end{figure}

\begin{table}[htb]
\caption[Table]{
Centroids (in MeV) of the single pion kinetic energy distributions for the 
$\pi^{+} \rightarrow \pi^{+}\pi^{-}$ channel ($\cal C$$_+$ and $\cal C$$_-$)
and for the $\pi^{+} \rightarrow \pi^{+}\pi^{+}$ channel ($\cal C$$_+$). 
The uncertainties ($\sigma$) in evaluating $\cal C$'s are reported in 
parenthesis.
}
\begin{center}
  \begin{tabular}{lcccc}                      \hline \hline 
    \multicolumn{1}{c}{Nucleus} & 
     \multicolumn{2}{c}{$\pi^{+} \rightarrow \pi^{+}\pi^{-}$ }  && 
      \multicolumn{1}{r}{$\pi^{+} \rightarrow \pi^{+}\pi^{+}$ }        \\ 
           & $\cal C$$_+$  & $\cal C$$_-$       && $\cal C$$_+$        \\ \hline
 $^{2}H$   &   50.7(0.4)   &      45.5(0.4)     &&     52.5(0.5)       \\ 
$^{12}C$   &   47.5(0.4)   &      40.8(0.4)     &&     48.4(0.5)       \\ 
$^{40}Ca$  &   48.9(0.5)   &      38.6(0.5)     &&     49.4(0.6)       \\ 
$^{208}Pb$ &   50.7(0.5)   &      37.2(0.5)     &&     49.3(0.7)       \\ \hline\hline
\end{tabular}
\end{center}
\end{table}

Fig. 14 shows the differential cross sections for the two 
$\pi^+ A \rightarrow \pi^+\pi^{\pm} A'$ reactions as a function of the 
kinetic energy of each final pion.
The weighted averages of the distributions are reported in Table 5.
$\cal C$$_+$ and $\cal C$$_-$ refer to the centroids of the distributions,
in MeV, for positive and negative pions, respectively. 
The $\cal C$$_+$'s display a weak $A-$dependence.
For the complex nuclei the 
mean energy is 49.0$\pm$1.0 MeV which is similar to the mean energy 
of 51.6$\pm$0.9 MeV for $^2H$.  This can be explained by observing that the 
nuclear Coulomb repulsion on final pions is in part compensated by the 
same repulsion on initial pions. In the $\pi^+\rightarrow\pi^+\pi^-$ 
channel, centroids follow a Coulomb-like behaviour, although the intensity 
is smaller than expected. In fact, $\cal C$$_-$ decreases by 2.2 MeV from 
$^{12}C$ to $^{40}Ca$, and by 1.4 MeV from $^{40}Ca$ to $^{208}Pb$, while a 
semi-classical treatment of the nuclear Coulomb attraction on point-like 
particles yields a decrease of 2.2 MeV and 4.6 MeV, respectively. 

In Fig. 14 the experimental
$T_{\pi}$ distributions are shown, along with results 
from $\pi^+ A \rightarrow \pi^+\pi^+ n [A-1]$ phase space simulations.
Except for $^2H$, the experimental distributions are well reproduced 
by the simulations, especially at higher energies, thus 
giving confidence in the reliability of the data. The result of the
simulations cannot be applied to negative pions, since the energy losses
of $\pi^{-}$'s due to Coulomb 
interactions is not exactly known.  The kinetic energy 
distributions are not compared to the results of any theoretical
predictions because none are 
available at this time. 

5.4 The $\pi$ angles

\begin{figure}[t]
 \centering
  \includegraphics*[angle=0,width=0.5\textwidth]{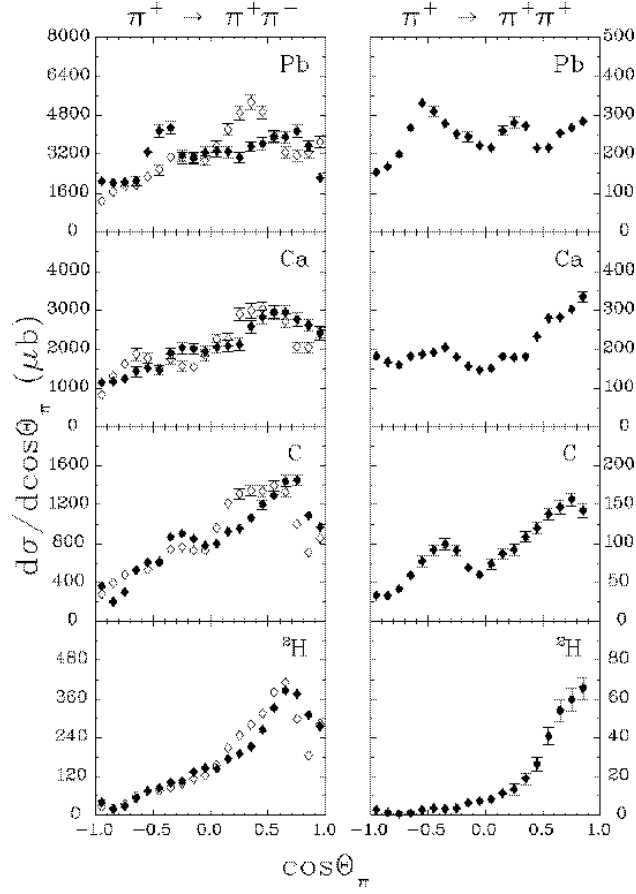}
  \setlength{\abovecaptionskip}{10pt}  
  \setlength{\belowcaptionskip}{15pt}  
  \caption{\footnotesize Distributions of cos$\Theta_{\pi}$ for the 
      $\pi^+ \rightarrow \pi^+\pi^-$ and $\pi^+ \rightarrow \pi^+\pi^+$ 
      reactions on $^{2}H$, $^{12}C$, $^{40}Ca$ and $^{208}Pb$. Positive 
      and negative pion angular distributions are in the laboratory frame,
      and are denoted with full and open diamonds, respectively.}
\end{figure}

In order to complete the discussion of the canonical observables, in Fig.
15 the cos$\Theta_{\pi}$ differential cross sections are reported, 
where pion angles are in the laboratory frame and cos$\Theta_{\pi}$=1
coincides with the direction of the incident pion. Positive and negative
pion distributions are denoted with full and open diamonds, respectively.
The interpretation of these distributions awaits 
microscopic $\pi 2\pi$ calculations, since phase space simulations cannot
describe the observed bump structures.

5.5 The $\cal C$$_{\pi\pi}^A$ ratio

In this paragraph the observable $\cal C$$_{\pi\pi}^A$ is presented 
in comparison with recent theoretical predictions. $\cal C$$_{\pi\pi}^A$ 
is defined as the composite ratio $\frac{M_{\pi\pi}^A}{\sigma_T^A} / 
\frac{M_{\pi\pi}^N}{\sigma_T^N}$, where $\sigma_T^A$ ($\sigma_T^N$) 
is the measured total cross section of the $\pi 2\pi$ process in nuclei 
(nucleon). This observable has the property of yielding the net effect 
of nuclear matter on the $(\pi\pi)_{I=J=0}$ interacting system regardless 
of the $\pi 2\pi$ reaction mechanism used to produce the pion pair
\cite{Bonutti:four}. Therefore, $\cal C$$_{\pi\pi}^A$ can be compared 
not only with the $\mathcal{M} 1$ and $\mathcal{M} 2$ predictions which
explicitly 
calculate both $M_{\pi\pi}^{Ca}$ and $M_{\pi\pi}^{^{2}H}$, but also with 
the theories described in $\mathcal{M} 3$ and $\mathcal{M} 4$ because they 
calculate the mass distribution of an interacting $(\pi\pi)_{I=J=0}$ 
system (i.e. $ImD_{\sigma}$) both in vacuum and in nuclear matter. Since 
the calculations are reported either in arbitrary units
\cite{Vicente:one,Rapp:one} or in units which are complex to scale
\cite{Hatsuda:one,Aouissat:two}, theoretical predictions are normalized 
to the experimental distributions at $M_{\pi\pi}$=350$\pm$10 MeV, where 
the $\cal C$$_{\pi\pi}^A$ distribution is flat.

For both reaction channels, the full\cite{Vicente:one} and dotted
\cite{Rapp:one} curves in Fig. 16 are obtained by simply dividing 
$M_{\pi\pi}^{Ca}$/ $M_{\pi\pi}^{^{2}H}$. It is worthwhile recalling 
that for \cite{Rapp:one} the option $\rho$=0.5$\rho_n$ is used while 
for \cite{Vicente:one} the mean density is $\rho$=0.24$\rho_n$.
Furthermore, for both approaches the underlying medium effect is the 
$P-$wave coupling of $\pi$'s to $p-h$ and $\Delta-h$ configurations, 
which accounts for the near-threshold enhancement. When applied to the 
$\cal C$$_{\pi\pi}^{Ca}$, both $\mathcal{M} 1$ and $\mathcal{M} 2$ predict 
the same result, which describes the behaviour of 
$\cal C$$_{++}^{Ca}$ fairly well
throughout most of the $M_{\pi\pi}$ energy range, but only reproduces
a small part of the near-threshold strength for $\cal C$$_{+-}^{Ca}$.

$\mathcal{M} 3$ \cite{Hatsuda:one} and $\mathcal{M} 4$ \cite{Aouissat:two} 
examine the medium modifications on the scalar-isoscalar meson, the 
$\sigma-$meson. Nuclear matter is assumed to partially restore chiral 
symmetry, and consequently $m_{\sigma}$ is assumed to vary with $\rho$.  
The variation is parametrized as $1-p\frac{\rho}{\rho_n}$, where $p$ is 
0.1$\leq p \leq$0.3 for $\mathcal{M} 3$, and 0.2$\leq p\leq$0.3 for 
$\mathcal{M} 4$, and for both is $\rho$=$\rho_n$. Both models are capable 
of yielding large strength near the $2m_\pi$ threshold, therefore the 
results for $\cal C$$_{+-}^A$ are compared for a common minimum value of 
the parameter $p$=0.2. In addition, the choice $\rho$=$\rho_n$ may result 
appropriate for $^{208}Pb$ but it is not for medium (i.e. $^{40}Ca$) and 
light (i.e. $^{12}C$) nuclei. In Fig. 16 the predictions of $\mathcal{M} 3$ 
and $\mathcal{M} 4$ are shown as the dash-dotted curves and dashed curves, 
respectively. $\mathcal{M} 4$ predicts a larger near-threshold strength, 
which is due to the combined contributions of the in-medium $P-$wave 
coupling of pions to $p-h$ and $\Delta-h$ configurations, and to the 
partial restoration of chiral symmetry in nuclear matter. These two 
models, however, are still too schematic for a conclusive comparison to 
the present data, and full theoretical calculations are called for.
\begin{figure}[t]
 \centering
  \includegraphics*[angle=90,width=0.7\textwidth]{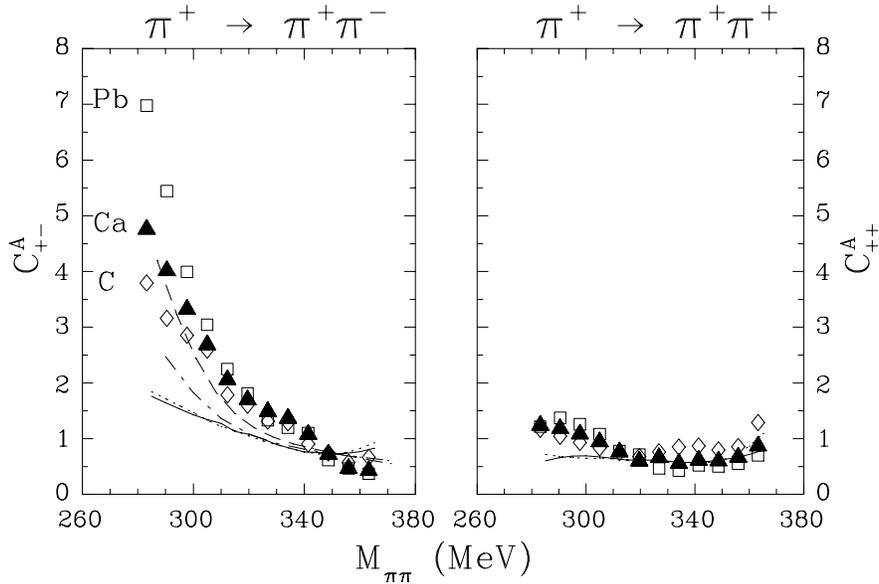}
  \setlength{\abovecaptionskip}{10pt}  
  \setlength{\belowcaptionskip}{15pt}  
  \caption{\footnotesize The composite ratios $\cal C$$_{\pi\pi}^A$
      for $^{12}C$, $^{40}Ca$ and $^{208}Pb$. The curves are taken from 
      \cite{Vicente:one} (full) and the model is described in 
      $\mathcal{M} 2$, \cite{Hatsuda:one} (dash-dotted) $\mathcal{M} 3$, 
      \cite{Rapp:one} (dotted) $\mathcal{M} 1$ and \cite{Aouissat:two} 
      (dashed) $\mathcal{M} 4$. Further details are reported in the text.}
\end{figure}

{\bf 6. Existing $\pi 2\pi$ results} \\

Novel $\pi 2\pi$ results \cite{Nefkens:one} were recently presented by
the Crystal Ball (CB) Collaboration at the AGS. The  $\pi^- A \rightarrow 
\pi^{\circ}\pi^{\circ} A'$ production reaction was studied on several nuclei 
$H$, $^{12}C$, $^{27}Al$ and $^{64}Cu$ at incident pion momenta (energies) 
of $p_{\pi^-}$=408 and 750 MeV/c ($T_{\pi^-}$=291.6 and 623.3 MeV). The
figures presented in Ref.\cite{Nefkens:one}, however, are
only for the results obtained at the higher momentum. 
In the near-threshold region the $\pi^+\pi^-$ and $\pi^{\circ}\pi^{\circ}$ 
channels can be directly compared since they share similar dynamical
aspects.  The $\pi^{\circ}\pi^{\circ}$ system cannot be in a $P-$state,
and $\pi^+\pi^-$ does not couple to $P-$waves (see Paragraph 5.1).
Also, the scattering amplitudes of the two reaction channels,
$\pi^-\rightarrow\pi^{\circ}\pi^{\circ}$ and $\pi^+ \rightarrow \pi^+\pi^-$,
have the two leading isospin amplitudes $T_{3,2}$ and $T_{1,0}$
in common, where the first (second) 
index is the total isospin (dipion isospin) \cite{Vicente:one}. The 
$M_{\pi^{\circ}\pi^{\circ}}^A / M_{\pi^{\circ}\pi^{\circ}}^C$ ratio in 
Ref.\cite{Nefkens:one} has marked similarities to the observable 
$\cal C$$_{\pi^+\pi^-}^A$.  Namely, the two ratios are flat at around 
400 MeV invariant mass but increase as $M_{\pi\pi}$ approaches the 
$2m_{\pi}$ threshold, and, near threshold, both ratios increase as $A$ 
increases.  Unlike the CHAOS data, the
invariant mass distributions in \cite{Nefkens:one} show
no evidence of a peak at the $2m_{\pi}$ threshold.  However,
the CB and CHAOS spectra do share relevant features: they both have
meager $M_{\pi\pi}$ strength near threshold for $H$ ($^2H$ for CHAOS), and 
show a dramatic increase of strength with increasing $A$.
As outlined earlier in the present work, the {\em 
CHAOS peak} may be attributed to the finite out-of-plane acceptance of 
the spectrometer, but the pion distributions are relatively unaffected
by nuclear distortions because of the low kinetic energies
of the two final pions \cite{Bonutti:two,Bonutti:four}. The mean kinetic 
energy of the CB $\pi^{\circ}$'s, however, may exceed 200 MeV, and at this 
energy pions may undergo large distortions while leaving the nucleus
\cite{Huffner:one}. Pions may be absorbed and the absorption rate depends on 
both the pion kinetic energy and its initial position inside a nucleus. In 
addition, pions scatter via the $\pi A\rightarrow\pi^{'}A$ reaction
and change direction.  This smears out the $\pi^{\circ}\pi^{\circ}$
opening angle distributions, and ultimately the $M_{\pi^{\circ}\pi^{\circ}}$ 
ones (see discussion in Paragraph 5.2). Such distortions are difficult to 
model since they depend on both the pion energy and the position inside the 
nucleus.  Thus, in order to study $\pi\pi$ medium modifications, it is
advisable to deal with low-energy final pions \cite{Bonutti:two,Bonutti:four}.

{\bf 7. Conclusions} \\

In this article the results of an exclusive measurement of the pion-production 
$\pi^+ A \rightarrow \pi^+\pi^{\pm} A'$ reactions on $^{2}H$, $^{12}C$, 
$^{40}Ca$, and $^{208}Pb$, at an incident pion energy $T_{\pi^+}$=283 MeV,
were presented. The primary interest was directed to the study of the 
$\pi\pi -$dynamics in nuclear matter. The reaction was initially examined on 
deuterium in order to understand the elementary pion-production mechanism,
then on nuclei in order to determine the effects of medium modifications
on the $\pi\pi -$system. Some of latter effects could be obtained by direct
comparison of the 
$\pi 2\pi$ distributions, since the data were taken under the same kinematical
conditions. The $\pi\pi$ properties were highlighted by means of $M_{\pi\pi}$, 
$cos\Theta_{\pi\pi}^{CM}$, $\Theta_{\pi\pi}$, etc. differential cross sections
as well as the composite $\cal C$$_{\pi\pi}^A$ ratios. 

$\cal C$$_{\pi\pi}^A$ was found to yield the net effect of nuclear matter 
on the $\pi\pi$ system regardless of the $\pi 2\pi$ reaction mechanism used 
to produce the pion pair. These distributions display a marked dependence 
on the charge state of the final pions. (i) The $\cal C$$_{\pi^+\pi^-}^A$ 
distributions peak at the $2m_{\pi}$ threshold and their yield increases as 
$A$ increases, thus indicating that pion pairs form a strongly interacting 
system.  Furthermore, the $\pi\pi$ system couples to $I=J=0$, 
the $\sigma -$meson channel. (ii) In the $\pi^+ \rightarrow \pi^+\pi^+$ 
channel, the $\cal C$$_{\pi\pi}^A$ distributions show little dependence
on either $A$ or $T$, thus indicating that nuclear matter only weakly
affects the $(\pi\pi)_{I,J=2,0}$ interaction. 

The $\cal C$$_{\pi^+\pi^-}^A$ observable was compared with theories
which included
the $(\pi\pi)_{I=J=0}$ in-medium modifications associated with the partial 
restoration of chiral symmetry in nuclear matter, and with model calculations
which only included standard many-body correlations, i.e. the $P-$wave 
coupling of $\pi's$ to $p-h$ and $\Delta-h$ configurations. It was found 
that both mechanisms are necessary to interpret the data, although chiral 
symmetry restoration yields the larger near-threshold contribution. If
this conclusion is correct, the $\pi 2\pi$ CHAOS data would provide an
example of a distinct $QCD$ effect in low-energy nuclear physics. 

Monte Carlo simulations of the $\pi^+ A \rightarrow \pi^+\pi^{\pm} N [A-1]$ 
reaction phase space proved useful in the interpretation of some of
the $\pi 2\pi$ 
data. In the case of $M_{\pi^+\pi^+}$, the distribution of
$\pi^+\pi^+$ pairs follows
phase space. In addition, simulations are able to describe 
the high-energy part of the distributions, which are sensitive to the nuclear 
Fermi momentum of the interacting $\pi^+ p[A-1] \rightarrow \pi^+\pi^+ 
n[A-1]'$ proton. The same conclusions apply to the $T_{\pi^+}$ distributions,
thus giving confidence in the reliability of the data. For the 
$M_{\pi^+\pi^-}$ distributions, the $\pi\pi$ dynamics overwhelms the dipion 
kinematics.  Unlike the phase space simulations,
the near-threshold $\pi^+\pi^-$ yield is 
suppressed in the elementary production reaction, $\pi^+$ $^2H \rightarrow 
\pi^+\pi^- pp$ in the present work, while, in the same energy range, medium 
modifications strongly enhance $M_{\pi^+\pi^-}$. Any interpretation of the 
shape of the $M_{\pi^+\pi^{\pm}}^A$ (or $\cal C$$_{\pi^+\pi^{\pm}}^A$) 
distributions should combine the effects of the restoration of chiral 
symmetry in nuclear matter, and standard many-body correlations. Such an 
approach would not require a high-density nuclear medium to obtain strong 
$\pi\pi$ medium modification in the proximity of the $2m_\pi$ threshold.

The results of the present measurement were compared with results from
the Crystal Ball Collaboration at the BNL, which are the only relevant
ones available. The two results show remarkable similarities. The 
difference in shape of the near-threshold $M_{\pi\pi}^A$ distributions 
can be ascribed to the limited CHAOS acceptance and to nuclear pion 
distortions in the case of the CB data. Most important,
however, is that the exclusive $\pi 2\pi$ CB measurement independently 
confirms that the nuclear medium strongly influences the $\pi\pi$ 
interaction in the $I=J=0$ channel.

The experiment was performed at the TRIUMF Meson Facility by using the 
CHAOS spectrometer.  The 
experimental method used in a previous measurement at TRIUMF and the 
results obtained were similar to those discussed in the present article. 
However, the first-generation equipment employed and the $\pi^+ A 
\rightarrow\pi^+\pi^- A'$ model used to interpret the results were able 
to disclose only part of the underlying physics.  

{\bf Acknowledgements} \\

The authors would like to acknowledge the support received from TRIUMF. 
The present work was made possible by grants from the Istituto Nazionale 
di Fisica Nucleare (INFN) of Italy, the National Science and Engineering 
Research Council (NSERC) of Canada, the Australian Research Council. The 
authors would also like to acknowledge useful discussions with Z. Aouissat, 
G. Chanfray, T. Hatsuda, E. Oset, R. Rapp, P. Schuck, M. Vicente-Vacas 
and J. Wambach.

\newpage
%
%
 
\end{document}